\documentclass[prd,preprintnumbers,nofootinbib]{revtex4} 
\usepackage{graphicx} 
\usepackage{amsmath}
\usepackage{amsfonts,amsbsy}
\usepackage{amssymb}
\usepackage{appendix}
\usepackage{array}
\usepackage{multirow}
\def\be{\begin{equation}}
\def\ee{\end{equation}}
\def\bea{\begin{eqnarray}}
\def\eea{\end{eqnarray}}

\def\eq#1{{Eq.~(\ref{#1})}}
\def\fig#1{{Fig.~\ref{#1}}}

\newlength{\dinwidth}
\newlength{\dinmargin}
\renewcommand{\vec}[1]{\boldsymbol{#1}}
\newcommand{\dif}{\mathrm{d}}

\newcommand{\chisq}{\chi^2/\mathrm{d.o.f.}}

\makeatother

\makeatother

\begin{document}

%\title{ Impact-parameter dependent saturation model and new combined HERA data }
\title{Analysis of  combined HERA data in the Impact-Parameter dependent Saturation model }

\author{Amir H. Rezaeian$^{1,2}$, Marat~Siddikov$^{1}$, Merijn Van de Klundert$^{3,4}$ and Raju Venugopalan$^{3}$}
\affiliation{
$^1$ Departamento de F\'\i sica, Universidad T\'ecnica
Federico Santa Mar\'\i a, Avda. Espa\~na 1680,
Casilla 110-V, Valparaiso, Chile\\
$^2$  Centro Cient\'\i fico Tecnol\'ogico de Valpara\'\i so (CCTVal), Universidad T\'ecnica
Federico Santa Mar\'\i a, Casilla 110-V, Valpara\'\i so, Chile\\
$^3$ Physics Department, Bldg. 510A, Brookhaven National Laboratory, Upton, NY 11973, USA\\
$^4$ Physics Department, University of Amsterdam, Science Park 904, 1098 XH Amsterdam, The Netherlands}
\date{\today}
\begin{abstract}
The Impact-Parameter dependent Saturation Model (IP-Sat) is a simple dipole model that incorporates key features of the physics of gluon saturation and matches smoothly to the perturbative QCD dipole expression at large $Q^2$ for a given $x$. It was previously shown that the model gives a good description of HERA data suggesting evidence for gluon saturation effects at small $x$. The model has also been applied to proton-proton and proton-nucleus collisions and provides the basis for the IP-Glasma model of initial conditions in heavy ion collisions. Here we present a reanalysis of available data in electron-proton collisions at small Bjorken-$x$, including the recently released combined data from the ZEUS and H1 collaborations. We first confront the model to the high precision combined data for the reduced cross-section and obtain its parameters. With these parameters fixed,  we compare model results to data for the structure function $F_2$, the longitudinal structure function $F_L$, the charm structure function $F_2^{c\bar{c}}$, exclusive vector meson ($J/\psi, \phi$ and $\rho$) production and Deeply Virtual Compton Scattering (DVCS). Excellent agreement is obtained for the processes considered at small $x$ in a wide range of $Q^2$. Our results strongly hint at universality of the IP-Sat dipole amplitude and the extracted impact-parameter distribution of the proton. They also provide a benchmark for further refinements in studies of QCD saturation at colliders.  
   
\end{abstract}
\maketitle

\section{Introduction}

Deep inelastic scattering (DIS)  and exclusive diffractive processes in electron-proton (ep) collisions such as exclusive vector meson production and deeply virtual compton scattering  (DVCS) at small Bjorken-$x$ are excellent probes of the high-energy limit of QCD, in particular the parton saturation regime \cite{sg,mv,cgc-review1}. An effective field theory describing this regime is the Color Glass Condensate (CGC)~\cite{mv,cgc-review1}. A key ingredient in particle production in the CGC approach is the universal dipole amplitude, the imaginary part of the quark-antiquark scattering amplitude on a proton or nuclear target. The evolution of the dipole amplitude with rapidity, given a suitable initial condition, is computed most generally by solving the Jalilian-Marian-Iancu-McLerran-Weigert-Leonidov-Kovner (JIMWLK)  hierarchy of equations~\cite{jimwlk}. In the large $N_c$ limit however, the coupled JIMWLK equations simplify to a closed form expression for the dipole amplitude known as the Balitsky-Kovchegov (BK) equation \cite{bk,bb}.

A common problem in QCD weak coupling approaches, to which the JIMWLK equations are not immune, is the impact parameter dependence of distributions; different final states will have varying degrees of sensitivity to large impact parameters where intrinsically non-perturbative physics may be important. The impact parameter dependence of the dipole amplitude in the JIMWLK and BK equations is cumbersome to implement, and is often in practice assumed to be uniform in impact parameter. A simple dipole model that incorporates the physics of saturation and models the impact parameter dependence of gluon distributions is the IP-Sat saturation model \cite{Bartels:2002cj,Kowalski:2003hm,watt2007}.  This model for the dipole amplitude, whose form can be derived at the classical level in the CGC~\cite{mv}, contains an eikonalized gluon distribution which satisfies DGLAP evolution while explicitly maintaining unitarity. It also matches smoothly to the high $Q^2$ perturbative QCD limit. The impact parameter dependence of the amplitude is simple to implement in the IP-Sat model and is easily generalized from DIS off protons to DIS off  nuclei~\cite{Kowalski:2007rw,Kowalski:2008sa}. It is perhaps the most attractive feature of this model and allows one to confront a large body of HERA data on exclusive diffractive processes which cannot otherwise be described simply in saturation models\footnote{Note that in this paper we only focus on the IP-Sat model and we do not consider the b-CGC model \cite{watt-bcgc} which is an alternative impact-parameter dependent saturation model  that has been applied to many reactions including diffractive processes \cite{watt2007,watt-bcgc}, proton-proton \cite{pp-LR}, proton-nucleus \cite{pa-R}, and nucleus-nucleus collisions \cite{aa-LR}. }.

A disadvantage of this model is that it does not have the same grounding in fundamental theory as possessed by the rcBK and JIMWLK frameworks. In the latter, the basis for QCD evolution is the BFKL equation, and higher order computations can be performed systematically in the CGC. The IP-Sat model attempts to approach the saturation boundary via DGLAP evolution; the eikonalization of the gluon distribution represents higher twist contributions 
that are becoming important at small $x$. However,  computations including higher twist evolution are difficult to perform and enormously complicated; indeed, this difficulty was what motivated the CGC effective field theory approach in the first place.

With these comments and the stated caveats, we will explore here how well the IP-Sat model fares with the latest combined data and attempt a 
detailed analysis of this data. The spirit of this computation is similar to the extraction of  parton distributions in the collinear factorization approach. 
However, in contrast to the collinear factorization approach, the number of free parameters in the dipole approach is considerably less\footnote{In fairness it must be noted that the collinear approaches attempt to describe nearly all available data. Here we consider only data for 
$x\leq 10^{-2}$.}; in the IP-Sat model we have only 4 free parameters that are varied.  Recently, the H1 and ZEUS collaborations have released new combined data for inclusive DIS~\cite{Aaron:2009aa,Abramowicz:1900rp} with unprecedentedly high precision. Because of the extremely small error bars of the new combined data set, it is vital to reconfront the IP-Sat dipole model with these data  to examine the effects of the tighter constraints on model parameters. Another advantage of the combined datasets is that the data are for the first time given in terms of the reduced cross-section $\sigma_r$,  which is unbiased towards any theoretical assumption in extracting the structure functions $F_2$ and $F_L$.

The main purpose of this paper is to reexamine the IP-Sat model in view of  recent precise data from HERA and to obtain its free parameters from a fit.  An important application of our results will be to make predictions for nuclei, in particular for a future Electron-Ion Collider (EIC)~\cite{Boer:2011fh,Lappi:2010dd,Toll:2012mb}. Another is to apply the improved results for the dipole amplitude to make comparisons to data in proton-proton~\cite{Tribedy:2010ab}, proton-nucleus~\cite{Tribedy:2011aa} and nucleus-nucleus collisions~\cite{Schenke:2012wb}. 
We aim the paper to be self-contained. In sections II and III, we provide a concise review of the formalism for calculating the total DIS cross-section, structure functions and exclusive diffractive processes in the color dipole approach. In section IV, we introduce the IP-Sat dipole model. In section V, we present our detailed numerical analysis, and the main results; these are then confronted with HERA data. Section VI contains a summary of the key results of this work.

%%%%%%%%%%%%%%%%%%%%%%%%%%%%
%%%%%%%%%%%%%%%%%%%%%%%%%%%%
\section{\bf  Total DIS cross-section and structure functions}
In the dipole picture, the scattering of the virtual photon $\gamma^{\star}$ on the proton at small $x$ has a distinct chronology in light front time. First, the $\gamma^{\star}$ fluctuates into a quark-antiquark
pair (the so-called $q\bar{q}$ dipole) of flavor $f$. It then interacts with the proton via gluon exchanges and
emissions; finally the $q\bar{q}$ pair recombines to form the final-state virtual photon. 
Because the lifetime of this $q\bar{q}$ dipole at small $x$ is much longer than its typical interaction time with target, the total deeply inelastic cross-section for a given $x$ and $Q^2$ can be written in the factorized form \cite{ni,al1}, 
\begin{equation}\label{gp}
  \sigma_{L,T}^{\gamma^*p}(Q^2,x) = 2\sum_f \int \int d^2b\, d^2r\,\int_0^1 dz\, |\Psi_{L,T}^{(f)}(r,z;Q^2)|^2 
  \,\mathcal{N}\left(x,r,b\right),
\end{equation}
where $\mathcal{N}\left(x,r,b\right)$ is the imaginary part of forward $q\bar{q}$ dipole-proton scattering amplitude with transverse dipole size $r$ and impact parameter $b$ (see section IV). The light cone wavefunction $\Psi_{L,T}^{(f)}$ for
$\gamma^{\star}$ fluctuations into $q\bar{q}$ is computable in QED with $L,T$ denoting the
longitudinal and transverse polarizations of the virtual photon:
\begin{subequations}\label{wfs}
\begin{eqnarray}
  |\Psi_{T}^{(f)}(r,z;Q^2)|^2 \, & = & \,
  \frac{\alpha_{EM} \, N_c}{2\, \pi^2} \, \sum\limits_f \, e_f^2 \, 
  \left\{ \epsilon_f^2 \, [K_1 (\epsilon_f\,r)]^2 \, [z^2 + (1 - z)^2] 
  + m_f^2 \, [K_0 (\epsilon_f\,r)]^2 \right\}, \ \label{wav}\\
  |\Psi_{L}^{(f)}(r,z;Q^2)|^2 \, & = & \, 
  \frac{\alpha_{EM} \, N_c}{2 \, \pi^2} \, \sum\limits_f e_f^2 \, 
  \left\{ 4 \, Q^2 z^2 (1 - z)^2 \, [K_0 (\epsilon_f\,r)]^2 \right\}, \label{wavL}
\end{eqnarray} 
\end{subequations}
with 
\begin{equation}
\epsilon_f^2 = z \, (1-z)\, Q^2 + m_f^2, \label{eps}
\end{equation}
where $z$ is the fraction of the light cone momentum of the virtual
photon carried by the quark, $m_f$ is the quark mass,  $\alpha_{EM}$ is the electromagnetic fine structure constant, $e_f$ is the electric charge of a quark with flavor $f$, and
$N_c$ denotes the number of colors. For the light quarks, the gluon density is evaluated at $x = x_{Bj}$ (Bjorken-x), while for charm quarks we take $x = x_{Bj} \, (1 + 4\,m_c^2/Q^2)$.  

The proton structure function $F_2$ and the longitudinal structure function $F_L$ can be written in terms of
$\gamma^{\star}p$ cross-section as
\begin{eqnarray}
F_2(Q^2,x) &=& \frac{Q^2}{4\pi^2\alpha_{EM}} 
\left[\sigma_L^{\gamma^*p}(Q^2,x)+\sigma_T^{\gamma^*p}(Q^2,x)\right],\label{f2}\\
F_L(Q^2,x) &=& \frac{Q^2}{4\pi^2\alpha_{EM}}\sigma_L^{\gamma^*p}(Q^2,x). \ \label{FL}
\end{eqnarray}
The contribution of the charm quark to the wave functions in Eqs.
(\ref{wfs}) feeds into Eqs.~(\ref{gp}) and (\ref{f2}) directly giving
the charm structure function $F_2^{c\bar{c}}$. The reduced cross-section $\sigma_{r}$ which is a directly measurable observable is expressed in terms of the inclusive $F_2$ and  $F_L$ to be  
\[
\sigma_{r}\left(x,y,Q^{2}\right)=F_{2}\left(x,Q^{2}\right)-\frac{y^{2}}{1+(1-y)^{2}}F_{L}\left(x,Q^{2}\right),
\]
where $y=Q^2/(sx)$ is the inelasticity variable and $\sqrt{s}$ denotes the center of mass energy in $ep$ collisions. In the above expression, we neglected the contribution of the $Z$ boson which is important only at very large $Q^2$. 
%%%%%%%%%%%%%%%%%%%%%%%%%%%%%%%%%%%
%%%%%%%%%%%%%%%%%%%%%%%%%%%%%%%%%%%

%%%%%%%%%%%%%%%%%%%%%%%%%%%%%%%%%%%
%%%%%%%%%%%%%%%%%%%%%%%%%%%%%%%%%%%
\section{\bf Deeply virtual compton scattering and exclusive diffractive processes}
In the dipole picture, similar to the case of the inclusive deeply inelastic scattering cross-section, the differential cross-section for the exclusive diffractive process $\gamma^*+p\to E+p$ with a final-state vector meson $E=J/\Psi, \phi,\rho$  or a real photon $E=\gamma$  in DVCS, can be written in terms of a convolution of the dipole amplitude and the overlap wave-functions of photon and the exclusive final-state particle \cite{watt2007},  
\begin{equation}
  \frac{\dif\sigma^{\gamma^* p\rightarrow Ep}_{T,L}}{\dif t} = \frac{1}{16\pi}\left\lvert\mathcal{A}^{\gamma^* p\rightarrow Ep}_{T,L}\right\rvert^2\;(1+\beta^2),
  \label{vm}
\end{equation}
with the scattering amplitude\footnote{An apparent factor of $4\pi$ difference with ~\cite{watt2007} is due to our different convention for 
$\Psi_{T,L}$.}
\begin{equation} \label{am-i}
  \mathcal{A}^{\gamma^* p\rightarrow Ep}_{T,L} = \mathrm{2i}\,\int\!\dif^2\vec{r}\int_0^1\!\dif{z}\int\!\dif^2\vec{b}\;(\Psi_{E}^{*}\Psi)_{T,L}\;\mathrm{e}^{-\mathrm{i}[\vec{b}-(1-z)\vec{r}]\cdot\vec{\Delta}}\mathcal{N}\left(x,r,b\right), 
\end{equation}
where $\vec{\Delta}^2=-t$ with $t$ being the squared momentum transfer. Note that the amplitude in \eq{am-i} is purely imaginary and obtained under assumption that the dipole-proton S-matrix is purely real. The factor $(1+\beta^2)$ in \eq{vm} takes into account the missing real part of amplitude in \eq{am-i} where $\beta$ is the ratio of the real to imaginary parts of the scattering amplitude, see e.g. \cite{watt2007,beta,mrt},
\begin{equation} \label{eq:beta}
  \beta = \tan\left(\frac{\pi\lambda}{2}\right), \quad\text{with}\quad \lambda \equiv \frac{\partial\ln\left(\mathcal{A}_{T,L}^{\gamma^* p\rightarrow Ep}\right)}{\partial\ln(1/x)}.
\end{equation}
For exclusive diffractive processes one should also incorporate the skewedness effect due to the fact that the gluons attached to the $q\bar{q}$ can carry different light-cone fractions $x,x^{\prime}$ of proton. At NLO level, in the limit that $x^{\prime}<<x<<1$, the skewedness effect  \cite{ske} can be accounted for by simply multiplying the gluon distribution $xg(x,\mu^2)$ by a factor $R_g$ defined via \cite{watt2007,mrt}, 

\begin{equation} \label{eq:Rg}
  R_g(\gamma) = \frac{2^{2\gamma+3}}{\sqrt{\pi}}\frac{\Gamma(\gamma+5/2)}{\Gamma(\gamma+4)}, \quad\text{with}\quad \gamma \equiv \frac{\partial\ln\left[xg(x,\mu^2)\right]}{\partial\ln(1/x)}.
\end{equation}
We note that there is uncertainty with regard to how one incorporates the skewedness correction at small $x$, and the 
factor $R_g$ should be regarded as a phenomenological estimate. Nevertheless, the gluon distribution is mainly determined from the reduced cross-section (or structure functions) alone; the choice of $R_g$ will only slightly affect the parametrization of the former.

In the case of the DVCS for the real photon production, only the transverse component of the overlap wavefunction contributes. Similarly to \eq{wav}, after summing over the quark helicities, we have
\begin{equation}
  (\Psi_\gamma^*\Psi)_{T}^f = \frac{N_c}{2\pi^2}\alpha_{\mathrm{em}}e_f^2\left\{\left[z^2+(1-z)^2\right]\epsilon_f K_1(\epsilon_f r) m_f K_1(m_f r)+ m_f^2 K_0(\epsilon_f r) K_0(m_f r)\right\},
  \label{w-dvcs}
\end{equation}
where one should sum over quark flavors $f=u,d,s,c$. There are several different prescription for modeling the vector meson wavefuctions. Following Refs.\,\cite{Kowalski:2003hm,watt2007,beta}, here we assume that in analogy to the DVCS, 
the vector meson Fock space is effectively dominated by the $q\bar{q}$ Fock component. The overlap between the photon and the vector meson wave functions is then expressed as \cite{beta} 
\begin{align}
  (\Psi_V^*\Psi)_{T} &= \hat{e}_f e\, \frac{N_c}{4\pi^2 z(1-z)} \,
  \left\{m_f^2 K_0(\epsilon_f r)\phi_T(r,z) - \left[z^2+(1-z)^2\right]\epsilon_f K_1(\epsilon_f r) \partial_r \phi_T(r,z)\right\},
  \label{eq:overt}
  \\
  (\Psi_V^*\Psi)_{L} &=  \, \hat{e}_f e \, \frac{N_c}{4\pi^2}\,
  2Qz(1-z)\,K_0(\epsilon_f r)\,
  \left[M_V\phi_L(r,z)+ \frac{m_f^2 - \nabla_r^2}{M_Vz(1-z)}
    \phi_L(r,z)\right],
  \label{eq:overl}
\end{align}
where $\nabla_r^2 \equiv (1/r)\partial_r + \partial_r^2$, $M_V$ is the meson mass and the effective charge is defined $\hat{e}_f=2/3$, $1/3$, or $1/\sqrt{2}$, for $J/\psi$, $\phi$, or $\rho$ mesons respectively. In the photon case, the scalar part $\phi$ is given by modified Bessel functions, whereas for vector mesons various quark models provide strong indication that a hadron at rest can be modeled by Gaussian fluctuations in transverse separation. 
 We use the so-called boosted Gaussian wave-functions \cite{beta} which were found to provide a better description of data \cite{watt2007}. The boosted Gaussian wavefunction has several advantages over other commonly used models, namely it is more self-consistent and has the proper short-distance limit of $z(1-z)$ as $m_f\to 0$, and more importantly it is fully boost-invariant. The boosted Gaussian wavefunction has the following simple form \cite{beta}, 
\begin{equation}
  \phi_{T,L}(r,z) = N_{T,L} z(1-z)
  \exp\left(-\frac{m_f^2 \mathcal{R}^2}{8z(1-z)} - \frac{2z(1-z)r^2}{\mathcal{R}^2} + \frac{m_f^2\mathcal{R}^2}{2}\right).
\end{equation}
where the parameters $ N_{T,L}$ and $\mathcal{R}$ are determined by imposing the normalization condition and via a fit to experimentally measured leptonic decay width of the vector meson for the longitudinally polarized case \cite{watt2007}. For the vector meson production, the gluon density is evaluated at $x=x_{Bj}\left(1+M^2_V/Q^2\right)$. 

We have followed closely here Refs.\,\cite{watt2007}, where several improvements were implemented in the treatment of exclusive diffractive processes compared to the original paper of Kowalski and Teaney \cite{Kowalski:2003hm}. These  include incorporating the contribution of the real part of the forward scattering amplitude and the skewedness effect due to the parton distributions being off-forward. We note also that in earlier studies, an important phase factor $\exp\left(i(1-z)\vec{r}\cdot\vec{\Delta}\right)$ arising from the non-forward wave-functions \cite{bbb} was neglected. As shown by Kowalski, Motyka and Watt \cite{watt2007}, this extra phase factor is very important and we have included it in our analysis. In \eq{am-i}, the phase factors $\exp\left(-i\vec{b}\cdot\vec{\Delta}\right)$  and $\exp\left(i(1-z)\vec{r}\cdot\vec{\Delta}\right)$ effectively takes into account the size of vector mesons in impact-parameter and dipole transverse-size space, respectively.  Even with these significant improvements, there is considerable room for further improvement in the study of off-forward gluon distributions at small $x$, especially in the region where higher twist effects are important.

%%%%%%%%%%%%%%%%%%%%%%%%%%%%%
%%%%%%%%%%%%%%%%%%%%%%%%%%%%%
\section{Impact-parameter dependent saturation dipole model: IP-Sat model}
The common ingredient of the cross-sections in DIS, exclusive diffractive vector meson production and  DVCS is the universal $q\bar{q}$ dipole-target amplitude. 
As seen in Eqs.\,(\ref{vm},\ref{am-i}), the impact-parameter dependence of the dipole amplitude is crucial for describing  exclusive diffractive processes. 
For the total cross-section, the effect of the impact-parameter dependence of the dipole amplitude is not especially important and the $b$-dependence can be effectively incorporated by treating it as
 a step function and adjusting the overall normalization. In this way, one can still find a good fit for the structure functions and total DIS cross-section. However, a consequence of a trivial $b$-dependence leads to a pronounced dip in the $t$-distribution of vector meson production at large $|t|$. This is not observed in data and can therefore be ruled out \cite{watt2007}. The choice of the impact-parameter profile of the dipole amplitude entails non-perturbative effects coming into play that are beyond the perturbative non-linear BK or JIMWLK equation. Both of these small $x$ evolution equations generate a power law Coulomb tail, which is not confining at large distances \cite{al,bk-c,ana}. 
A simple $b$-dependence for the dipole amplitude is obtained by combining the Glauber-Mueller form \cite{Kowalski:2003hm,watt2007,al1}  of the amplitude
%\begin{align*}
\bea
\mathcal{N}\left(x,r,b\right)  &=&\left(1-\exp\left(-\frac{\pi^{2}r^{2}}{2N_{c}}\alpha_{s}\left(\mu^{2}\right)xg\left(x,\mu^{2}\right)T_{G}(b)\right)\right)\,, \label{ip-sat} 
\eea
with a Gaussian impact parameter profile
\bea
T_{G}(b)&=& \frac{1}{2\pi B_G}\exp\left(-b^2/2B_G\right) \label{ip-b} \, ,\
\eea
where $B_G$ is a dimensionful scale which, in some models, corresponds to the spatial string tension in QCD. In Eq.~(\ref{ip-sat}), $xg\left(x,\mu_{0}^{2}\right)$ is the gluon density evolved up to  the scale $\mu$ with LO DGLAP gluon evolution (neglecting its coupling to quarks).  
 Note that in the original IP-Sat fit, the number of flavors was taken to be 3 \cite{Kowalski:2003hm,watt2007}. However, since the parameter $B_G$ will be fixed with experimental data for exclusive $J/\Psi$ production and because we also want to describe the charm structure function, we take $N_f=4$. We then take the corresponding one loop running-coupling value of $\alpha_s$ with $\Lambda_{\text{QCD}}=0.156$ GeV fixed by the experimentally measured value of $\alpha_s$ at the $Z^0$ mass. The contribution from bottom quarks is neglected. As in the original IP-Sat model, the scale $\mu^2$ is related to the dipole transverse size by
\begin{equation}
 \mu^{2}=C/r^{2}+\mu_{0}^{2}, \label{c}
\end{equation}
and the initial gluon distribution at the scale $\mu_0^2$ is taken to be  
\begin{equation}
 xg\left(x,\mu_{0}^{2}\right) =A_{g}\,x^{-\lambda_{g}}(1-x)^{5.6} \label{g}.
\end{equation}
Following Refs.\,\cite{Kowalski:2003hm,watt2007},  the parameter $C$ is set fixed\footnote{The parameter $C$ is correlated with other parameters of the model and its value cannot be uniquely determined via a fit. } to $C=4$. Thus the parameters $A_{g},\lambda_{g}, \mu_{0}^{2}$ and $B_G$ are the only free parameters of our model which will be fixed by a fit to the reduced cross-section. At large values of $M^2_V+Q^2$, we are in the color transparency regime and the main contribution to \eq{vm} comes from small dipole sizes. Therefore the $t$-distribution at small dipole sizes can be approximately determined by the Fourier transform of $T_{G}(b)$,
\begin{equation}
 \frac{\dif\sigma^{\gamma^* p\rightarrow Ep}_{T,L}}{\dif t}\approx e^{-B_G|t|}, \label{bd}
\end{equation}
which is fully supported by the experimental data (see Sec. V). We will extract the values of $B_G$ from the $t$-distribution of exclusive vector meson and DVCS data. 

%%%%%%%%%%%%%%%%%%%%%%%%
\begin{figure}[t]       
                               \includegraphics[width=6.20 cm] {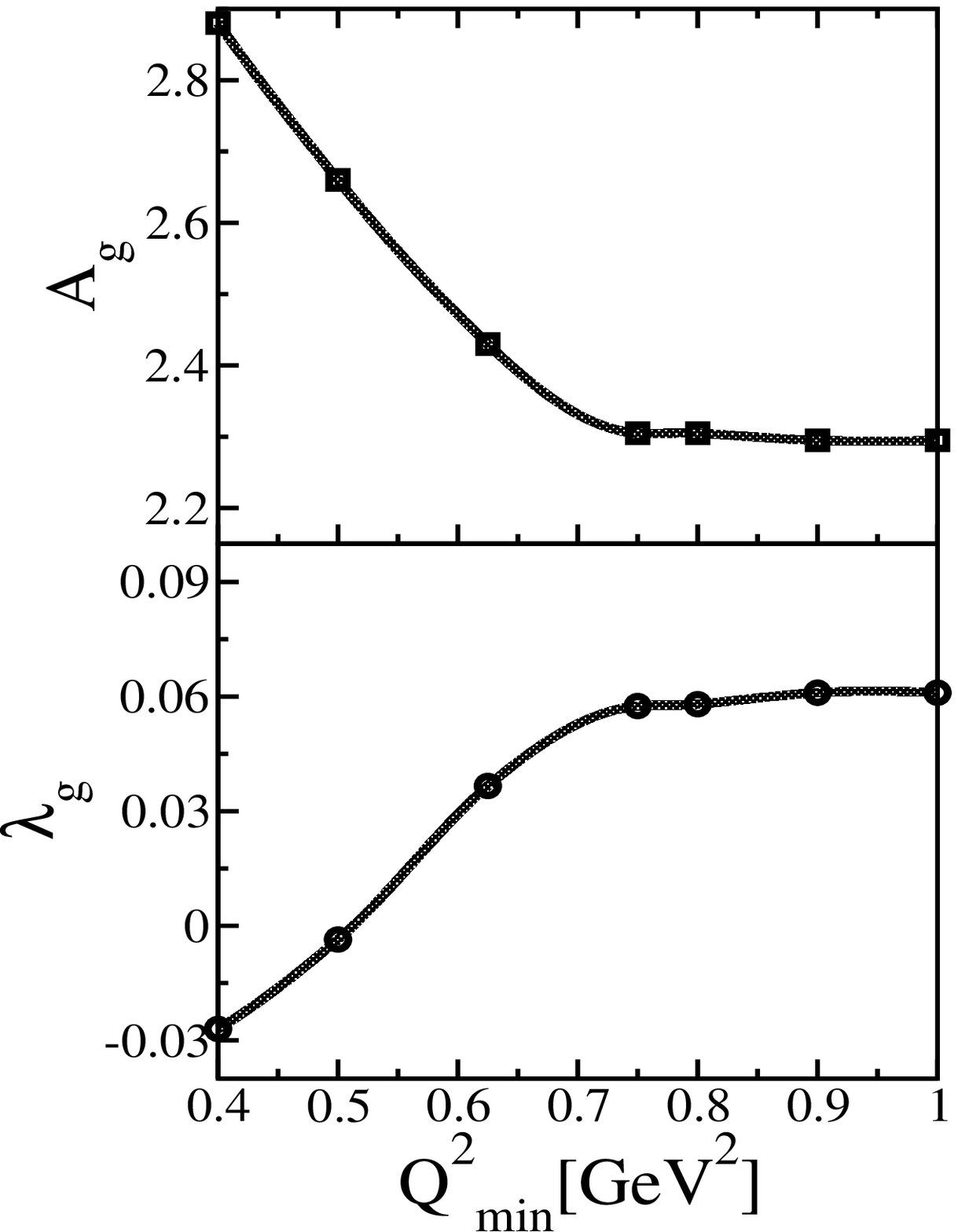}       
                                \includegraphics[width=6.20 cm] {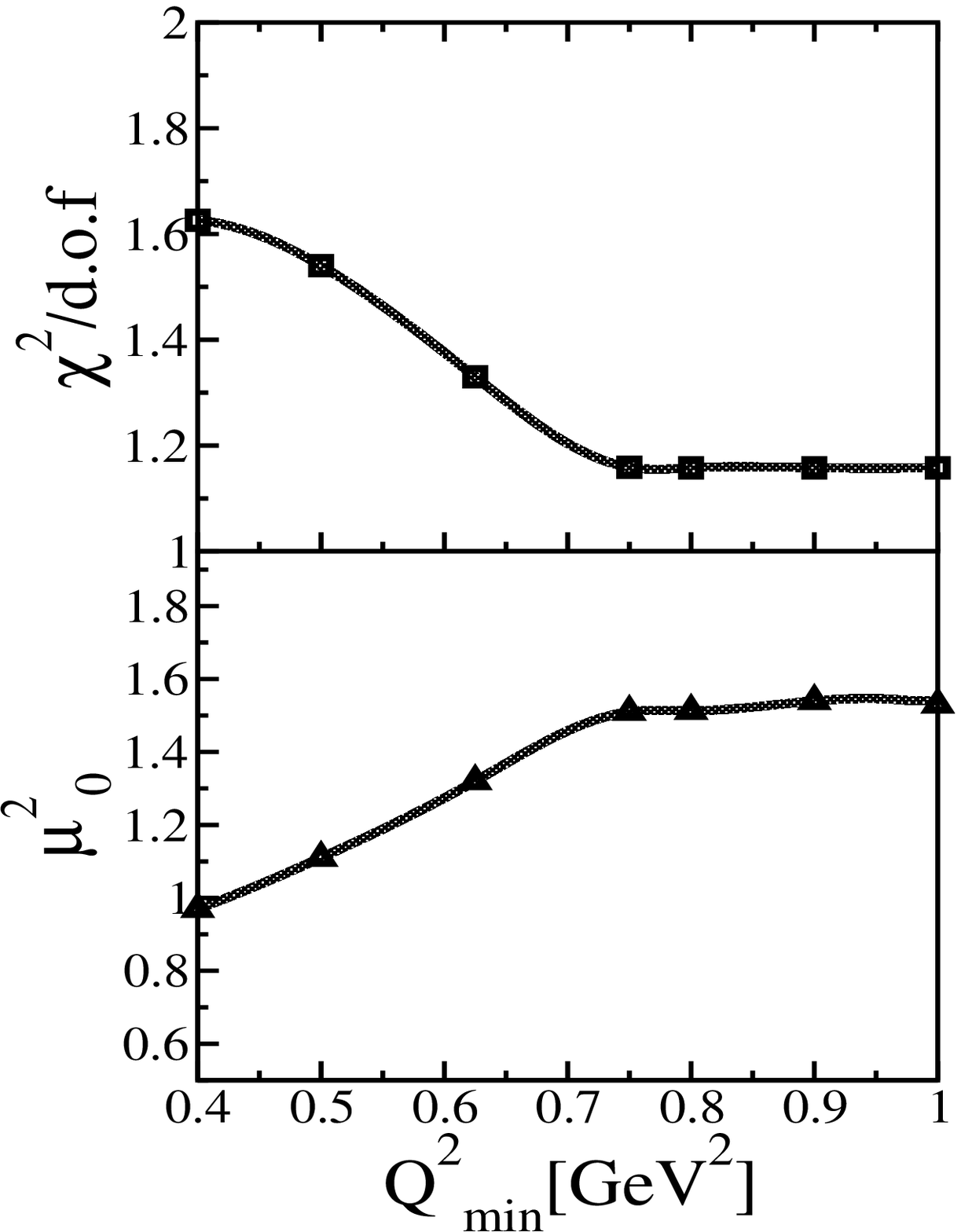}                                     
\caption{ The parameters of the IP-Sat model, $A_g, \lambda_g, \mu_0^2$ and the corresponding $\chi^2/d.o.f $ obtained from the fit as a function of lower virtuality cut $Q^2_{\text{min}}$ in the data bin selection.}
\label{f-chi1}           
\end{figure}
\begin{figure}[]     
\includegraphics[width=10 cm] {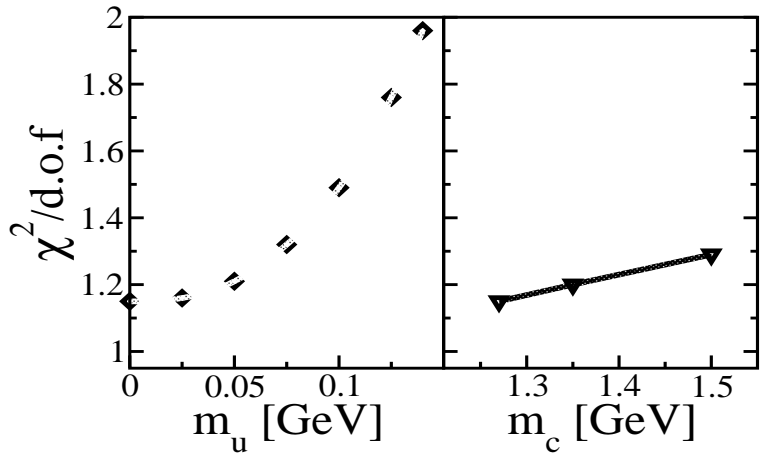}       
 \caption{ The $\chi^2/d.o.f $  of  the fit as a function of quark masses. In the left panel, we take $m_c=1.27$ GeV and vary the light quark mass while in the right panel, we take $m_u=0$ and vary the charm quark mass. }
\label{f-chi2}           
\end{figure}      
%%%%%%%%%%%%%%%%%%%%%%% 
%%%%%%%%%%%%%%%%%%%%%%%%%%%%%%%%%
\section{Numerical results and discussion}
We include in our fit the recently released data for the reduced cross-section $\sigma_r$ from the combined analysis of the H1 and ZEUS collaborations \cite{Aaron:2009aa}. In earlier analyses for obtaining a fit for IP-Sat model \cite{Kowalski:2003hm,watt2007}, H1 data was not included in the $\chi^2$ calculation in order to avoid introducing a normalization parameter between old H1 and ZEUS data. The new combined data from H1 and ZEUS collaborations are free from this problem and extremely precise. Unfortunately, experimental data for structure functions from other experiments such as muon DIS data from E665 \cite{Adams:1996gu} and NMC~\cite{Arneodo:1996qe} have rather larger error bars and will spoil the high quality data points of $\sigma_r$ from the combined H1 and ZEUS collaborations, even though including those data will improve the $\chi^2$ of the fit\footnote{We checked that indeed these fixed target data improve $\chi^2$ about $20\div 30\%$.}. Therefore we discard the experimental data for the $F_2$ structure function from E665 and NMC collaborations. We should emphasize further that our fit to extract the free parameters of the model is performed for the reduced inclusive DIS cross-section alone. With the extracted parameters, we then confront model results for $F_2$,$F_L$ and $F_{2}^{c\bar{c}}$ with the HERA data.  

In the calculation of $\chi^{2}$, the statistical and systematic experimental  uncertainties are added in quadrature.  
Since new combined data points have error bars as small as $\sim1\%$,
in extracting $\chi^{2}$, we have to evaluate all the theoretical observables with at least one or two orders of magnitude higher precision.
This causes the minimization algorithms to be very slow, a problem which can be overcome by implementing a parallelization of the numerical code. 

%%%%%%%%%%%%%%%%%%%%%%%%%%%%%%%%%
\begin{table}
  \centering
  \begin{tabular}{cccc|ccc|c}
    \hline\hline
    Data & $B_G/\mathrm{GeV}^2$ & $m_{u,d,s}$/GeV & $m_c$/GeV & $\mu_0^2/\mathrm{GeV}^2$ & $A_g$ & $\lambda_g$ & $\chisq$ \\ \hline
     $\sigma_r$ & 4 & $\approx 0$ & $1.27$ & $1.51$ & $2.308 $ & 0.058& $298.89/259 =1.15$ \\ \hline
      $\sigma_r$ & 4 & $\approx 0$ & $1.4$ & $1.428$ & $2.373 $ & 0.052& $316.61/259 =1.22$ \\ \hline\hline
  \end{tabular}
  \caption{Parameters of the initial gluon distribution in the IP-Sat model determined from fits to the reduced cross-section $\sigma_r$  from the combined H1 and ZEUS data in neutral current unpolarized $e^{\pm}p$ scattering  \cite{Aaron:2009aa} for data in the range $Q^2\in [0.75, 650]\,\text{GeV}^2$ and $x\le 0.01$. Results are shown for two fixed values of the charm quark masses. }
  \label{t-1}
\end{table}
%%%%%%%%%%%%%%%
\begin{figure}[t]       
\includegraphics[width=0.50\textwidth,clip]{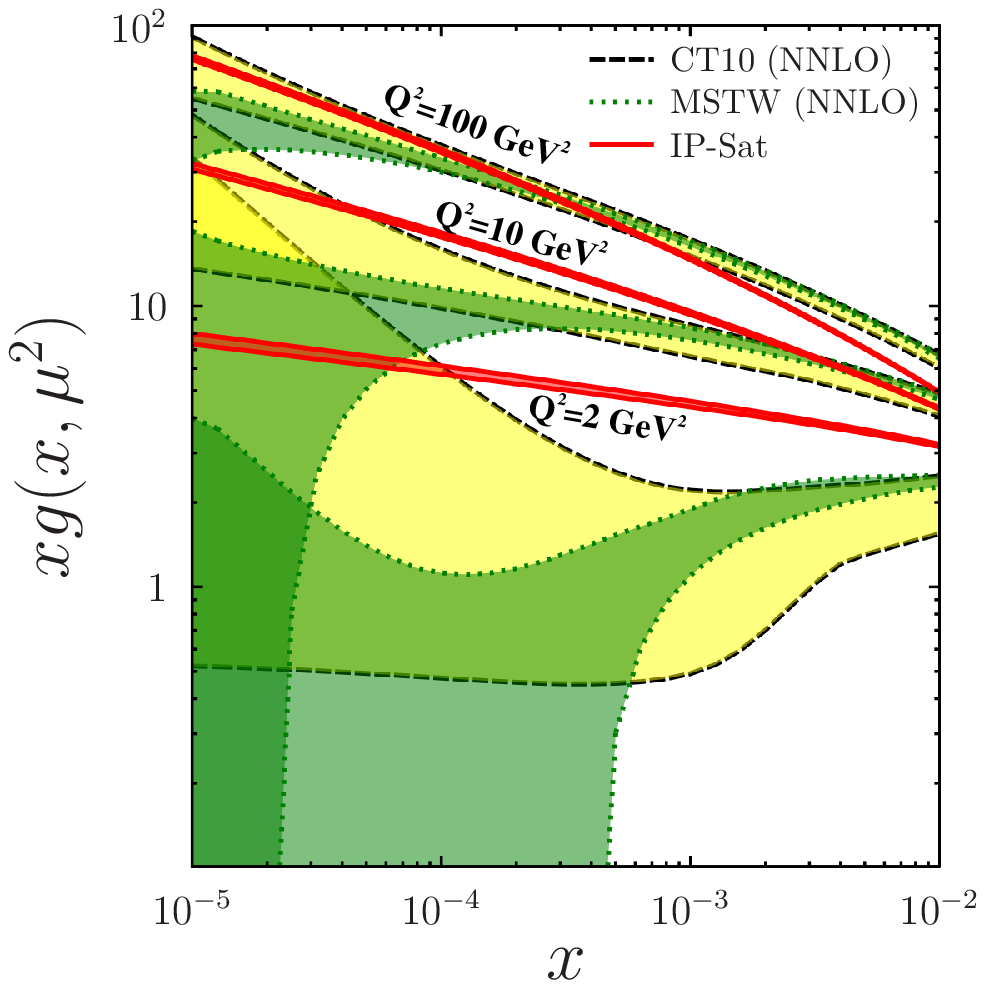}   
\includegraphics[width=0.49\textwidth,clip]{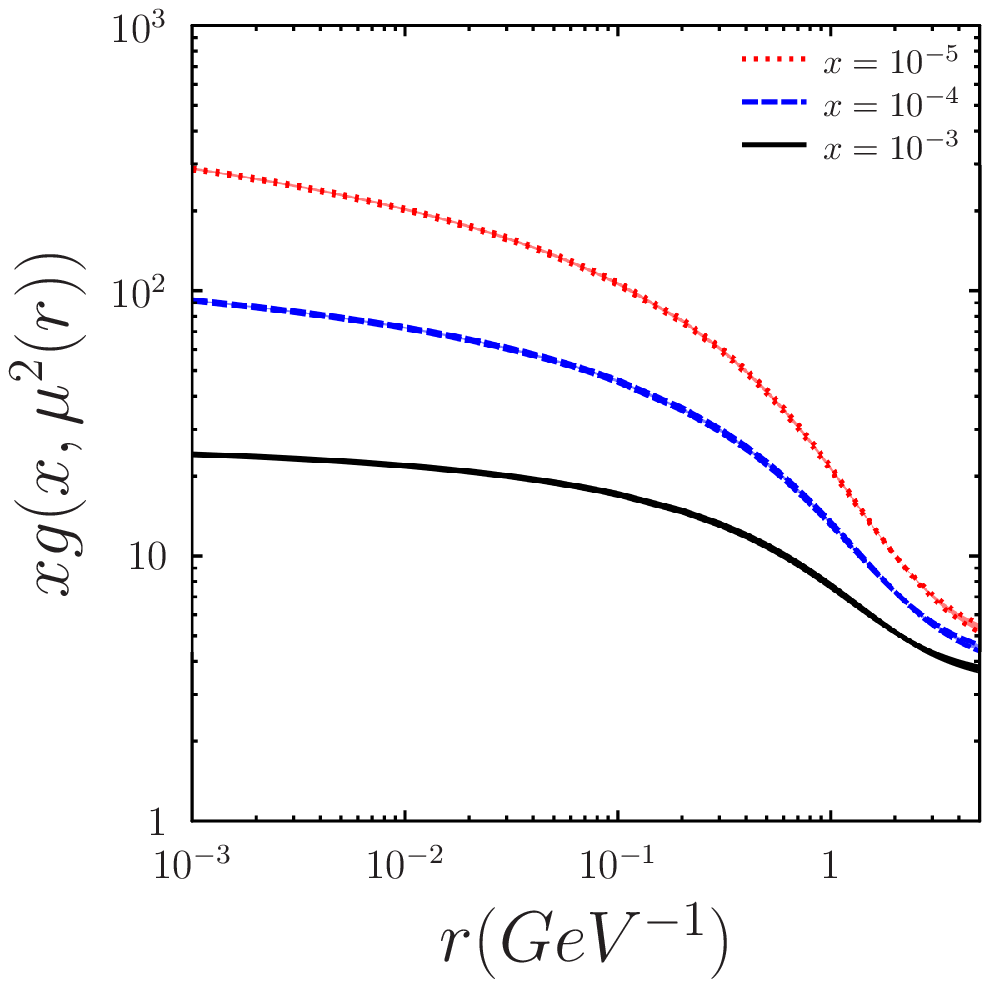}     
\caption{ Left: The gluon structure function as a function of $x$ for various fixed values of virtuality $Q^2$ extracted in the dipole saturation model (IP-Sat),  CT10 (NNLO) \cite{Lai:2010vv} and MSTW 2008 (NNLO) \cite{mstw}.  
The corresponding theoretical uncertainties are represented with bands between solid, dashed and dotted lines for IP-Sat, CT10 and MSTW, respectively. Right: The gluon structure function $xg\left(x, \mu^2(r)\right)$ as a function of dipole transverse size $r$ for various fixed values of $x$.}
\label{f-g}           
\end{figure}     
 %%%%%%%%%%%%%%%
%%%%%%%%%%%%%%%
\begin{figure}[t]       
\includegraphics[width=0.49\textwidth,clip]{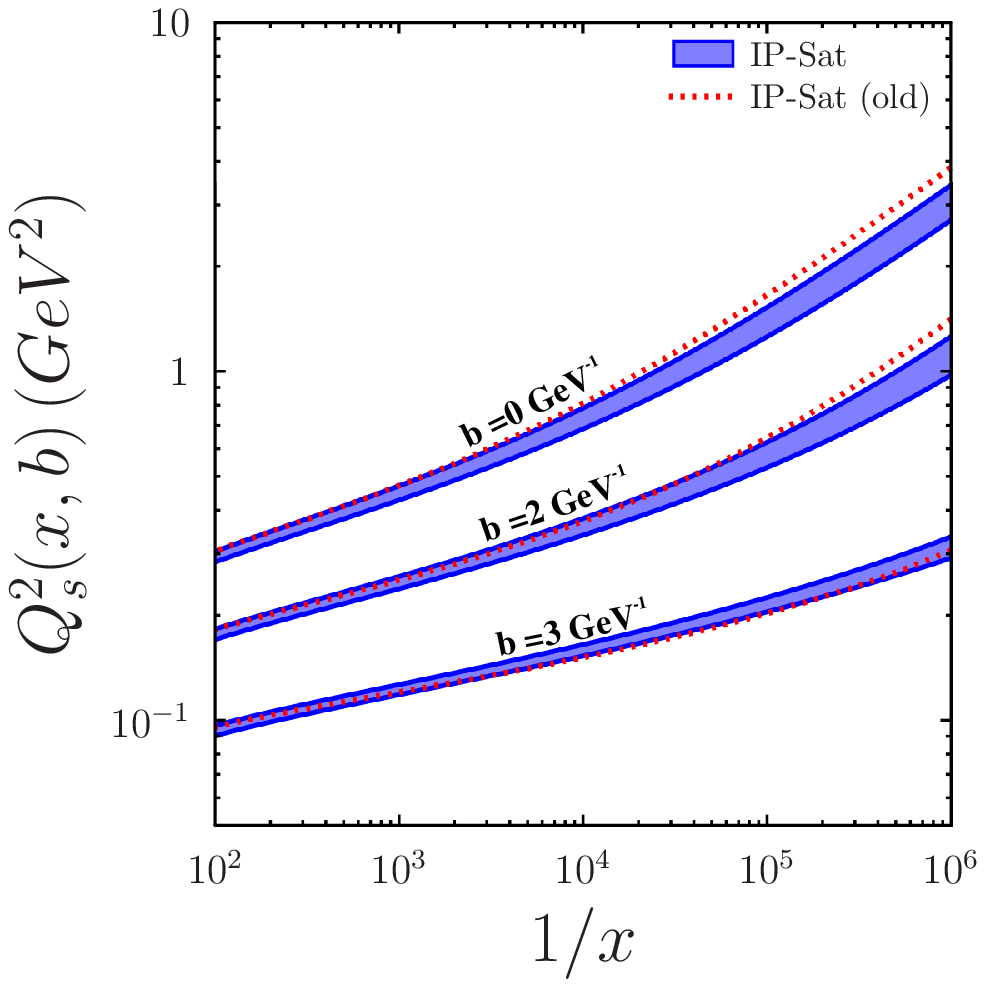}   
\includegraphics[width=0.49\textwidth,clip]{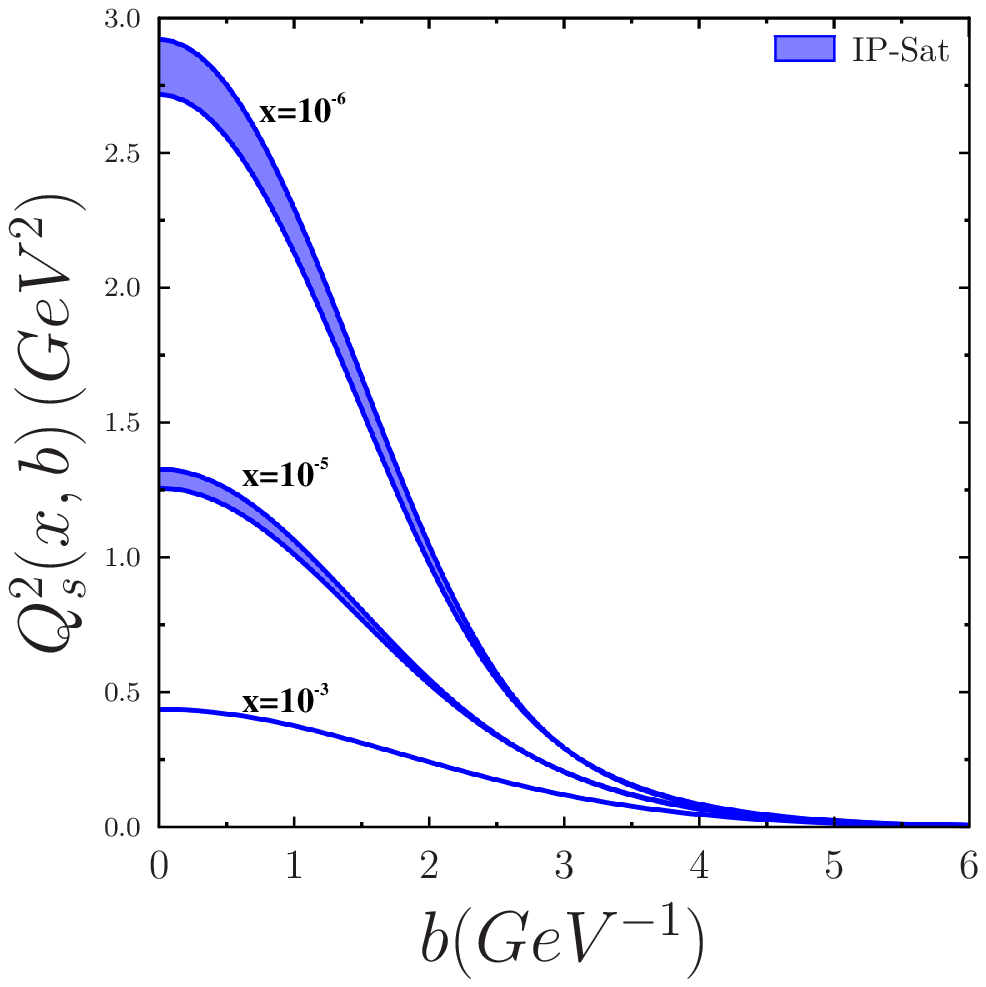}
 \caption{Left: The saturation scale extracted from the IP-Sat model with the parameter sets given in table \ref{t-1} and the old parameterization set from Ref.\,\cite{watt2007} labeled by IP-Sat (old) as a function of $1/x$ at various impact-parameter $b$. Right: The saturation scale as a function of the impact-parameter $b$ for various fixed values of $x$. In both panels, the lower and upper curves in the band correspond to the results obtained with the parameter sets given in table \ref{t-1} with charm mass $m_c=1.27, 1.4$ GeV, respectively. }
\label{f-g2}           
\end{figure}     
%%%%%%%%%%%%%%
%%%%%%%%%%%%%%

As noted previously, we have altogether $4$ free parameters which we fix via a fit to experimental data. Following Refs.\,\cite{Kowalski:2003hm,watt2007} , we do not allow the parameter $B_G$ to vary in addition to other parameters in the $\chi^2$  minimization algorithm, but we adjust it iteratively in order to obtain a good description of the $t$-dependence of exclusive diffractive processes.  We only consider data within the kinematic bin with $x\le 10^{-2}$ and $Q^2_{\text{min}}<Q^2[\text{GeV}^2]<650$. Although changing the upper limit cut on $Q^2$ will not greatly influence  our fit, imposing a lower cut on $Q^2_{\text{min}}$ significantly improves the fit. Note that in different studies, different values for a lower cutoff of the virtuality have been taken \cite{Kowalski:2003hm,watt2007,rest,KR}. Obviously, our model which is based on weak coupling dynamics is unreliable at very low virtualities on the order of or below the confinement scale. Moreover, the DGLAP evolution starts at the scale $\mu_0^2$, therefore at $Q^2<\mu_0^2$ we are sensitive to the parametrization of the gluon distribution at the initial scale.  In \fig{f-chi1}, we show the sensitivity of our parameters $A_g, \lambda_g$, and $\mu_0^2$ obtained from a fit with different data bin selections corresponding to  different lower cutoffs $Q^2_{\text{min}}$. We also show the corresponding $\chi^2/d.o.f$ as a function of $Q^2_{\text{min}}$. There are two main features in \fig{f-chi1} which guide us to choose the preferred value of $Q^2_{\text{min}}$. Namely, $\chi^2/d.o.f$  tends to decrease by increasing $Q^2_{\text{min}}$ and then flattens at about $Q^2_{\text{min}}\approx 0.75\,\text{GeV}^2$. Moreover, the values of all parameters  obtained from the fit become stable and saturate at $Q^2_{\text{min}}\approx 0.75\,\text{GeV}^2$. Therefore in the final fit shown in table \ref{t-1} we take HERA combined data for $\sigma_r$ within $Q^2\in [0.75, 650]\,\text{GeV}^2$ and $x\le 0.01$. Altogether we include 263 experimental data points for $\sigma_r$ from the combined H1 and ZEUS data in neutral current unpolarized $e^{\pm}p$ scattering  \cite{Aaron:2009aa}.  
\begin{figure}
  \centering
  \includegraphics[width=0.5\textwidth,clip]{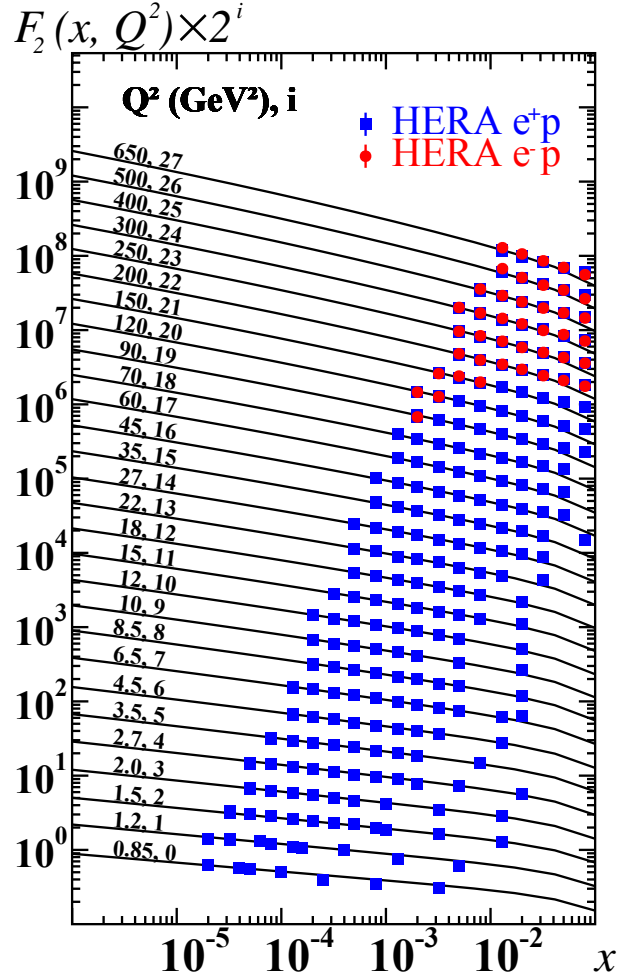}
  \caption{Results for the structure function $F_2(x,Q^2)$ as a function of $x$ for various values of $Q^2$. In order to separate data for each $Q^2$ from the others, the data and model results represented by the lines are multiplied by a factor $2^i$, with $i$ given on plot. We used the parameters set of the IP-Sat model  given in table \ref{t-1} with $m_c=1.27$ GeV. The experimental data are from combined H1 and ZEUS collaborations \cite{Aaron:2009aa}.  }
  \label{f-f2}
\end{figure}

In \fig{f-chi2}, we show the effect of light and charm quark masses in our fit. Remarkably, the data prefer very light current quark masses of order a few MeV. This seems to be natural given the fact that the saturation scale here is large relative to $\Lambda_{\rm QCD}$; our fit should therefore be less sensitive to non-perturbative infrared dynamics. To show more clearly the effect of light quark masses in our fit, we allow the  light quark mass $m_u$ ($m_u=m_d=m_s$) to vary as a free parameter and obtain its preferred value from the $\chi^2$ minimization. We find $m_u\approx 10^{-4}$ GeV from the fit.  It is seen however in \fig{f-chi2} (left panel) that the results of  the fit and minimization is stable with $m_u$ of order or smaller than few MeV.  In earlier analyses of the IP-Sat model, heavier quark masses of $m_u=50, 140$ MeV were required for a good fit \cite{Kowalski:2003hm,watt2007}, in sharp contrast to our conclusions from the new combined H1+ZEUS data. 

The charm mass dependence of our fit is shown in \fig{f-chi2} (right panel). It is seen that the minimization is more sensitive to light quark masses than charm mass. This is understandable given that charm data will not greatly influence our fit even if we include $\sigma_r^{c\bar{c}}$ data in our $\chi^2$ calculation. In table \ref{t-1}, we show our fits with two different charm masses $m_c=1.27$ and $1.4$ GeV. Note that the parameters of the IP-Sat model obtained here from the new combined data  are different from the earlier studies. If we take the old IP-Sat parameters with the same quark masses as in Ref.\,\cite{watt2007} and confront it with new data within $Q^2\in [0.75, 650]\,\text{GeV}^2$, we obtain $\chi^2/d.o.f=825.86/259=3.19$ which is not a good fit according to the hypothesis-testing criterion. The earlier IP-Sat fits should be therefore superseded by the fit given in table \ref{t-1}. 

In \fig{f-g} (left panel), we compare the gluon structure function  at various fixed values of virtuality $Q^2$ obtained from the dipole saturation model (IP-Sat) and the leading twist collinear factorization approach with NNLO DGLAP evolution,  namely CT10 \cite{Lai:2010vv} and MSTW 2008 \cite{mstw}. The bands for CT10 and MSTW correspond to uncertainties in obtaining a fit from global data analysis, while in the IP-Sat model the uncertainties are mainly due to our freedom to choose different values for the charm quark mass in the range $m_c=1.27 \div 1.4$ GeV. At large virtualities, we are in the color transparency region, the saturation effects become irrelevant and our approach approximately matches the standard perturbative formalism. The small differences seen at high virtualities are mainly due to the fact that we used LO DGLAP evolution without including quark degrees of freedom, while quark evolution contributions were included in the perturbative leading twist results shown in \fig{f-g}. At low virtualities and low $x$, we are in the saturation regime and we observe our gluon distributions to be significantly different from those obtained from the leading twist perturbative computations and significantly more stable than these, especially the MSTW fits. One may conclude that the higher twist contributions significantly influence the 
extraction of the gluon distribution in this regime.  In \fig{f-g} (right panel), we show the gluon structure function $xg\left(x, \mu^2(r)\right)$ in the IP-Sat model as a function of dipole transverse-size $r$ for various fixed values of $x$. Note that large dipole transverse-size corresponds to low virtuality via \eq{c}; consequently, the gluon distribution is reduced.

Following Refs.\,\cite{Kowalski:2003hm,watt2007}, we define the saturation scale $Q_s^2=2/r_s^2$, with $r_s$ being the saturation radius, as a scale where the dipole scattering amplitude has a value $\mathcal{N}(x,r_s,b)=\left(1-\exp(-1/2)\right)=0.4$. In \fig{f-g2} (left), we compare the saturation scale extracted from the IP-Sat model with old parametrization of Ref.\,\cite{watt2007} and the parametrization obtained in this paper as a function of $1/x$ at various impact-parameter $b$. It is seen that the saturation scale extracted from old and the new combined data from HERA are consistent with each other.  In \fig{f-g2} (right), we show the saturation scale as a function of the impact-parameter $b$ for different fixed values of $x$ obtained with parameter sets given in table \ref{t-1}. The band in \fig{f-g2} shows the theoretical uncertainties associated with our freedom to choose different values for the charm quark mass within $1.27\div 1.4$ GeV. It is generally seen that the saturation scale plotted as a function of  $1/x$ grows faster for more central collisions ($b\approx 0$). Moreover, the saturation scale at different impact-parameter can be significantly different, as large as one order of magnitude. This non-trivial behavior clearly show the importance of the impact-parameter dependence of the saturation scale.

\begin{figure}
  \centering
  \includegraphics[angle=270, width=1\textwidth,clip]{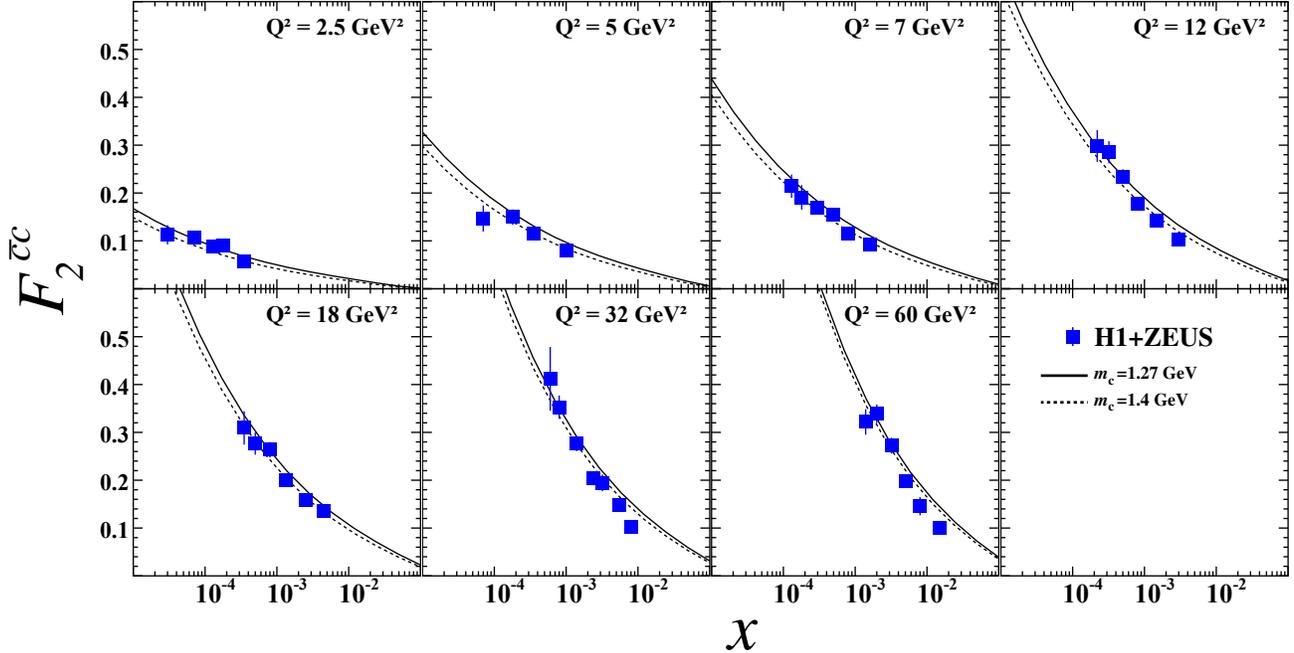}
  \caption{IP-Sat model results for the charm structure function $F_2^{c\bar{c}}(x,Q^2)$ as a function of $x$ for various values of $Q^2$ compared to HERA data. The solid and dashed lines are obtained with two different parameter sets given in table \ref{t-1} corresponding to charm mass $m_c=1.27, 1.4$ GeV respectively.  The experimental data points are from the recently released combined  data sets from the H1 and ZEUS collaborations \cite{Abramowicz:1900rp} assuming that $\sigma_r^{c\bar{c}}\approx F_2^{c\bar{c}}$ (see text). }
  \label{f-f2c}
\end{figure}

\begin{figure}
  \centering
  \includegraphics[width=0.7\textwidth,clip]{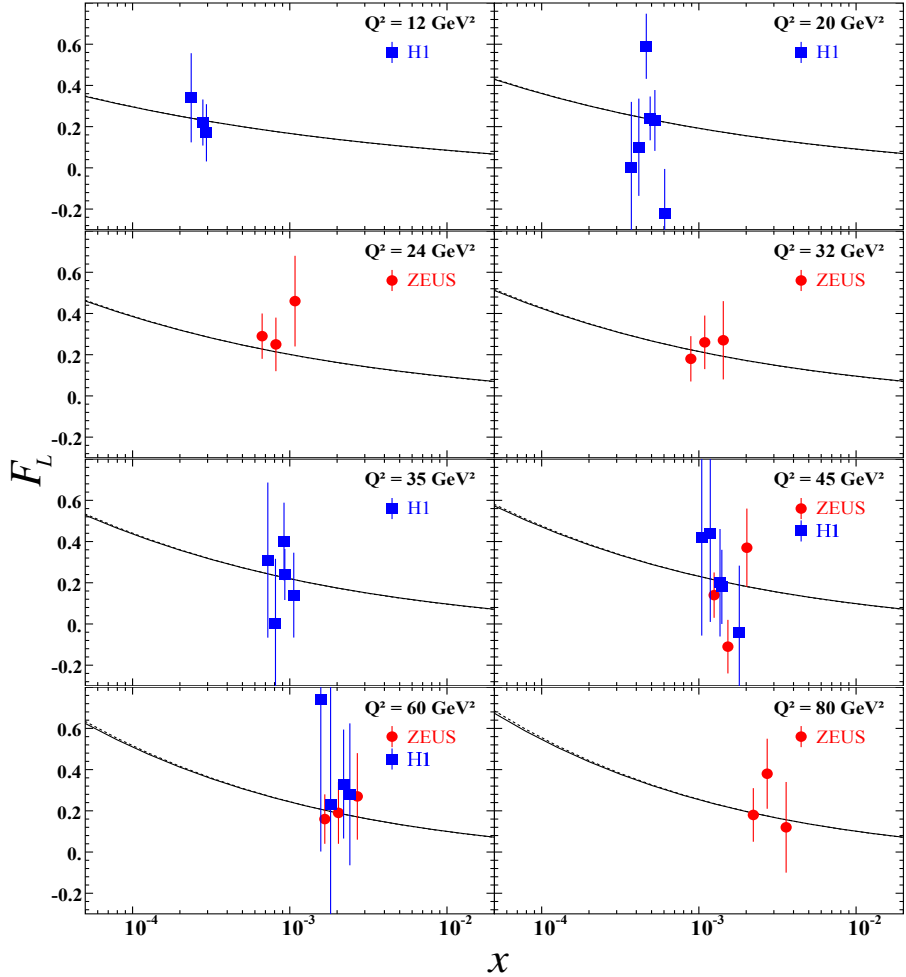}
  \caption{IP-Sat model results for the longitudinal structure function  $F_L(x,Q^2)$ as a function of $x$ for various values of $Q^2$ compared to HERA data.
We used the parameters set of the IP-Sat model  given in table \ref{t-1} with $m_c=1.27$ GeV. The experimental data are from \cite{Aaron:2008ad,Chekanov:2009na}. }
  \label{f-fl}
\end{figure}

With the parameters given in table \ref{t-1}, extracted from the $\chi$-squared fit to the reduced inclusive DIS cross-section, we now compute the structure functions $F_2(x,Q^2)$, the longitudinal structure function $F_L(x,Q^2)$ and the charm structure function $F_2^{c\bar{c}}(x,Q^2)$ in the IP-Sat model (using Eqs.\,(\ref{f2},\ref{FL},\ref{ip-sat})) and compare to the combined HERA data sets for the same. As emphasized previously, these $F_2$, $F_L$ and $F_2^{c\bar{c}}$ experimental data were not included in our fit and are therefore a prediction of the model based on the fit to the reduced inclusive DIS cross-section alone. It is seen in Figs.\,(\ref{f-f2},\ref{f-f2c},\ref{f-fl}) that the IP-Sat model results are in good agreement with structure functions data for a wide range of kinematics for $Q^2\in [0.75, 650]\,\text{GeV}^2$ and $x\le 0.01$. For the $F_2$ and $F_2^{c\bar{c}}$ we confronted model results with new H1+ZEUS combined data while for $F_L(x,Q^2)$  we compared with the existing data from HERA. 

The agreement of the IP-Sat model with data is most striking for the case of structure functions $F_2$ and $F_2^{c\bar{c}}$ where the experimental data are more precise. This agreement can also be taken as a non-trivial consistency check of the model. For the structure functions $F_2$, the agreement with data extends to $x > 0.01$ and only begins to fail closer to $x=0.1$. In Fig.\,(\ref{f-f2c}), we show the $F_2^{c\bar{c}}$ data from the H1+ZEUS combined analysis assuming that the contribution of $F_L^{c\bar{c}}$ to the reduced cross-section, originating from the exchange of longitudinally polarized photons, is negligible in the kinematic range considered here. This may give rise to less than a few per cent error to the data (not shown in Fig.\,(\ref{f-f2c})). It is seen in Fig.\,(\ref{f-fl}) that there are some irregularities in the experimental data points for  $F_L(x,Q^2)$ \cite{Aaron:2008ad} which await more precise measurements.  In Figs.\,(\ref{f-f2},\ref{f-f2c},\ref{f-fl}) we extend our model results beyond the kinematics of existing data, as predictions for future DIS experiments.

%%%%%%%%%%%%%%

%%%%%%%%%%%%%%%%%%%%%%%%%%%%

\begin{figure}
  %\centering
  \includegraphics[width=0.5\textwidth,clip]{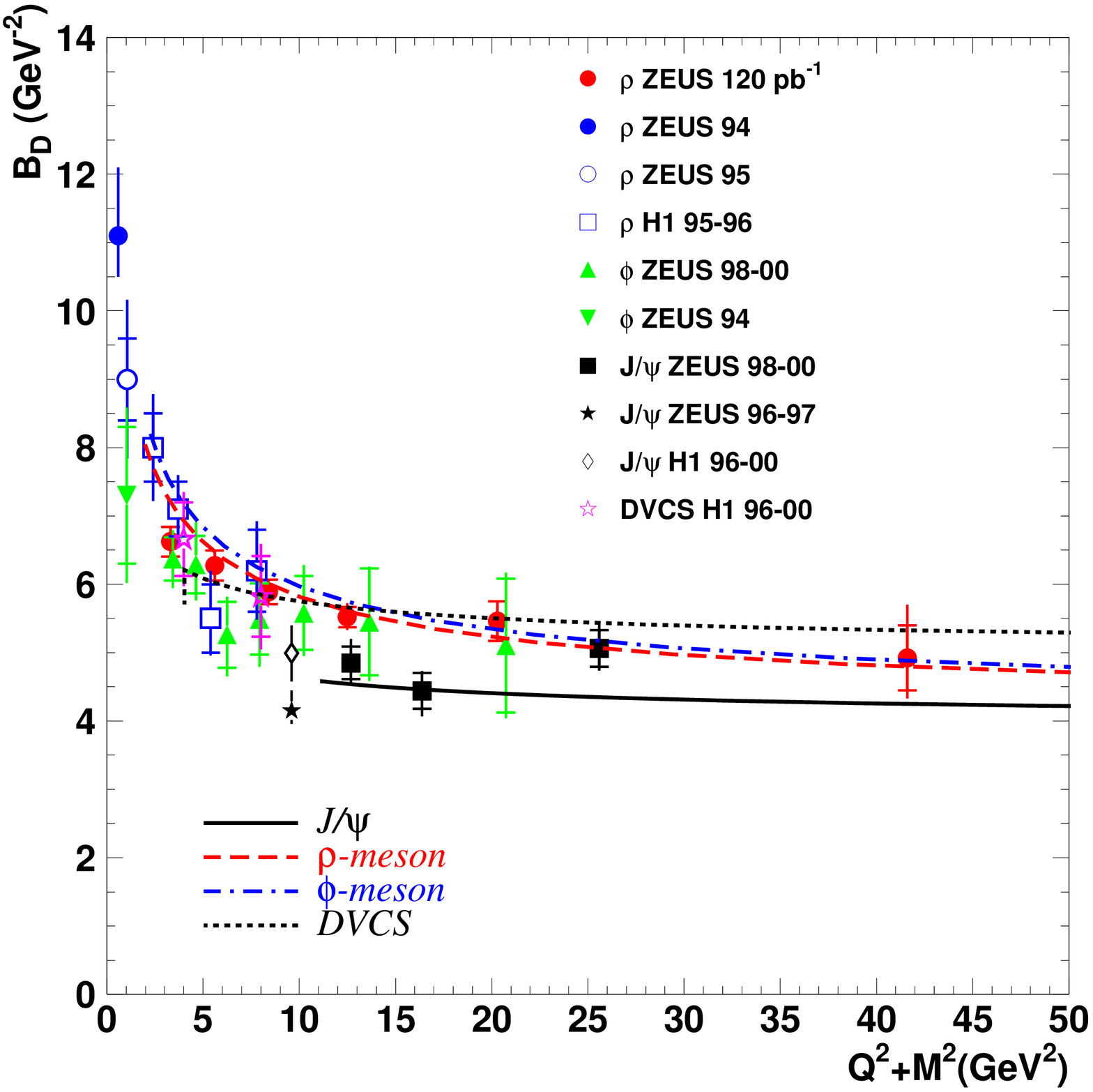}
  \caption{ A compilation of the value of the slope $B_D$ of $t$-distribution of exclusive vector-meson electroproduction and DVCS processes, as a function of $Q^2+M^2_V$.
The compilation of experimental data is from \cite{Chekanov:2007zr}. }
  \label{f-bd}
\end{figure}

We now turn our attention to results for exclusive diffractive processes at HERA. We first focus on the $t$-distribution of  exclusive vector meson production and Deeply Virtual Compton Scattering (DVCS). We fixed the width of proton impact-parameter profile $B_G$ in \eq{ip-b} via a fit to the slope of the $t$-distribution of the $J/\Psi$ mesons and we found
$B_G=4\,\text{GeV}^{-2}$. This value is in accordance with Ref.\,\cite{watt2007} since we used the same data set to fix the value of $B_G$.   In \fig{f-bd} we show a compilation of extracted values of the slope $B_D$ from a fit of the form $d\sigma/d|t| \propto e^{-B_D t}$ for exclusive vector-mesons electroproduction and DVCS. Note that\footnote{In general, $B_D\neq B_G$ because the 
Fourier transform of the $b$-distribution in the IP-Sat exclusive VM amplitude is not a simple exponential in $|t|$.} for very large $Q^2+M^2_V$ or small dipole-size, we have $B_D\approx B_G$, see \eq{bd}.  As it is seen in \fig{f-bd}, the experimental errors for the values of $B_D$ are rather large. This leads to some uncertainties in extracting the value of the parameter $B_G$. We estimated that the uncertainties of the value of $B_G$ is about $0.4\,\text{GeV}^{-2}$.

In \fig{f-bd}, only the line labeled by  $J/\Psi$ is a fit, the other lines corresponding to light mesons and the DVCS are results from the model. 
The extracted values of $B_D$ from theory at the same $Q^2+M^2_V$ are larger for lighter vector mesons than those for $J/\Psi$ in accordance with data. This is simply because due to the convolution of the photon wavefunction with the vector meson wavefunction, the typical dipole sizes for light and heavy mesons are different and mainly controlled by $1/\epsilon_f$ with $\epsilon_f$ given in \eq{eps}. Therefore, at a fixed virtuality, the typical dipole size is bigger for lighter vector meson and consequently the validity of the asymptotic expression in \eq{bd}  is postponed to a higher virtuality. The recent data from ZEUS for $\rho^0$ meson production \cite{Chekanov:2007zr} which extend to higher $Q^2$ (shown also in \fig{f-bd}), in accordance with the underlying assumptions of our model, indicate that indeed at large $Q^2+M^2_V$  the value of $B_D$  tends to saturate to a universal value mainly determined by the interaction area and shape of proton. 

\begin{figure}
  %\centering
  \includegraphics[angle=270,width=0.49\textwidth,clip]{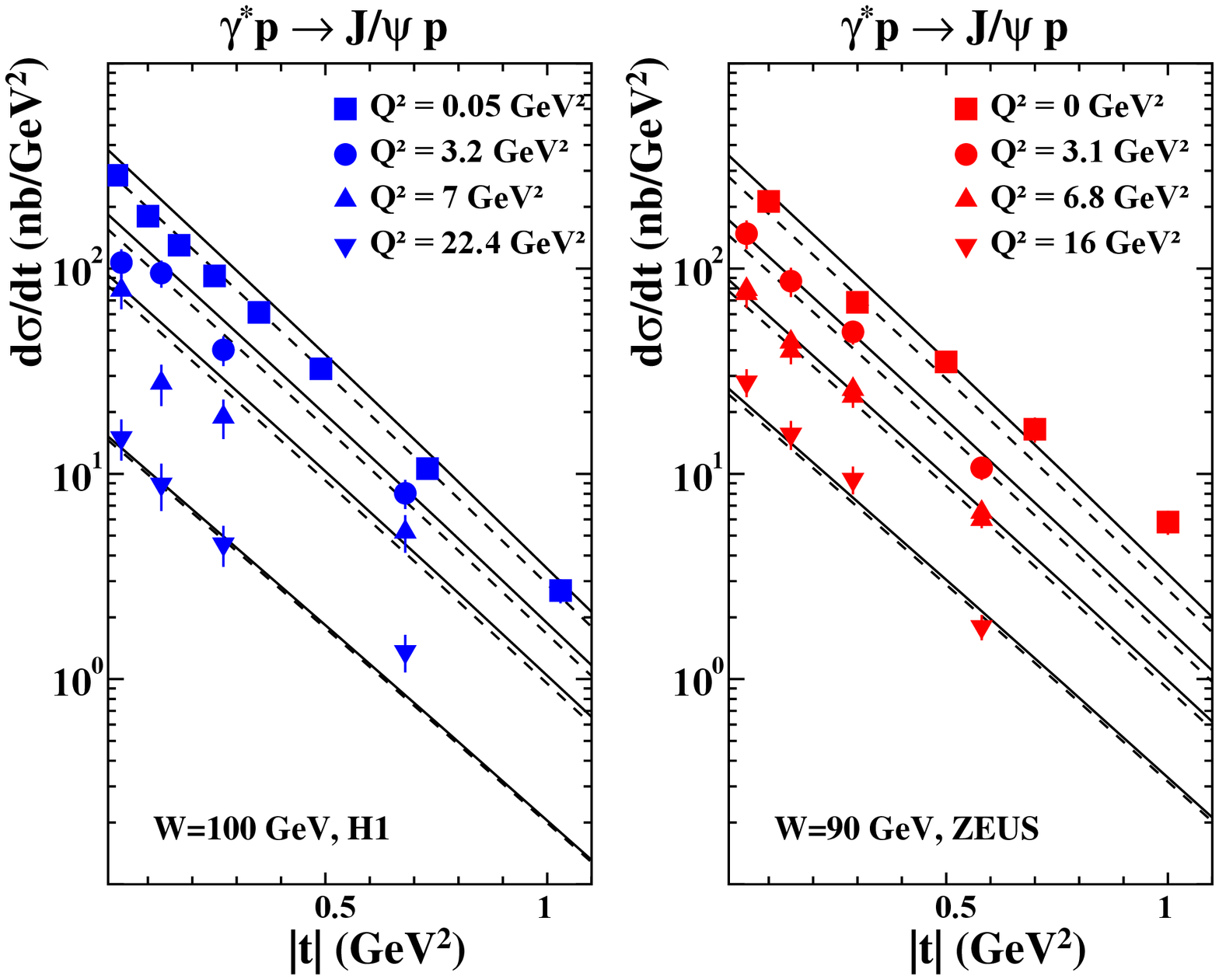}
\includegraphics[angle=270,width=0.49\textwidth,clip]{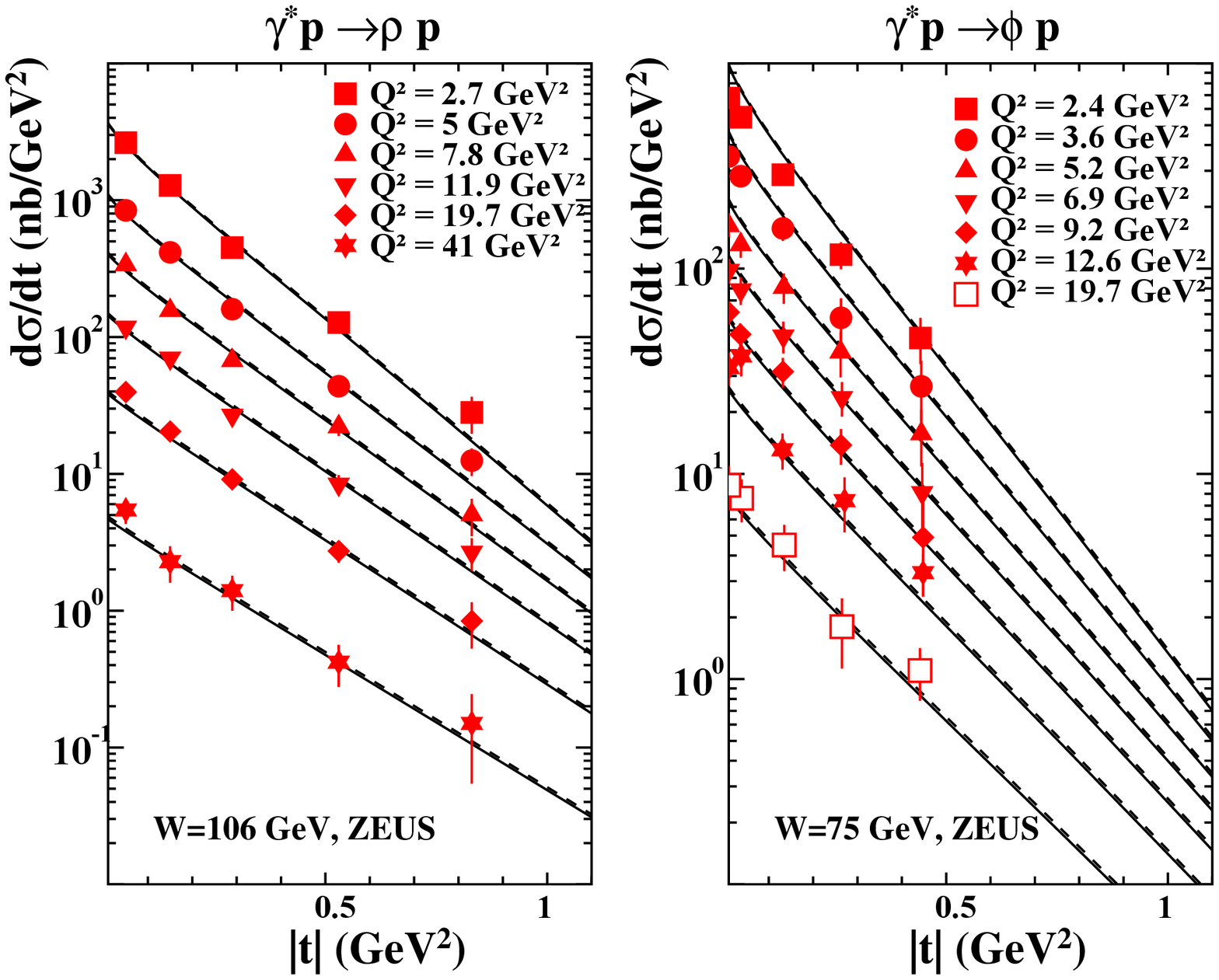}
  \caption{Differential vector meson cross-sections for $J/\Psi$, $\rho$ and  $\phi$ as a function of $|t|$. Data for a given $W$ with varying $Q^2$ are compared to results from the IP-Sat model using two parameter sets given in table \ref{t-1} with $m_c=1.27$ GeV (solid line) and $m_c=1.4$ GeV (dashed line).  The data are from the H1 and ZEUS collaborations \cite{Chekanov:2002xi,Chekanov:2004mw,Aktas:2005xu,Chekanov:2005cqa,Aaron:2009xp,Chekanov:2007zr}.  }
  \label{f-vt}
\end{figure}

\begin{figure}
  %\centering
  \includegraphics[angle=270,width=0.49\textwidth,clip]{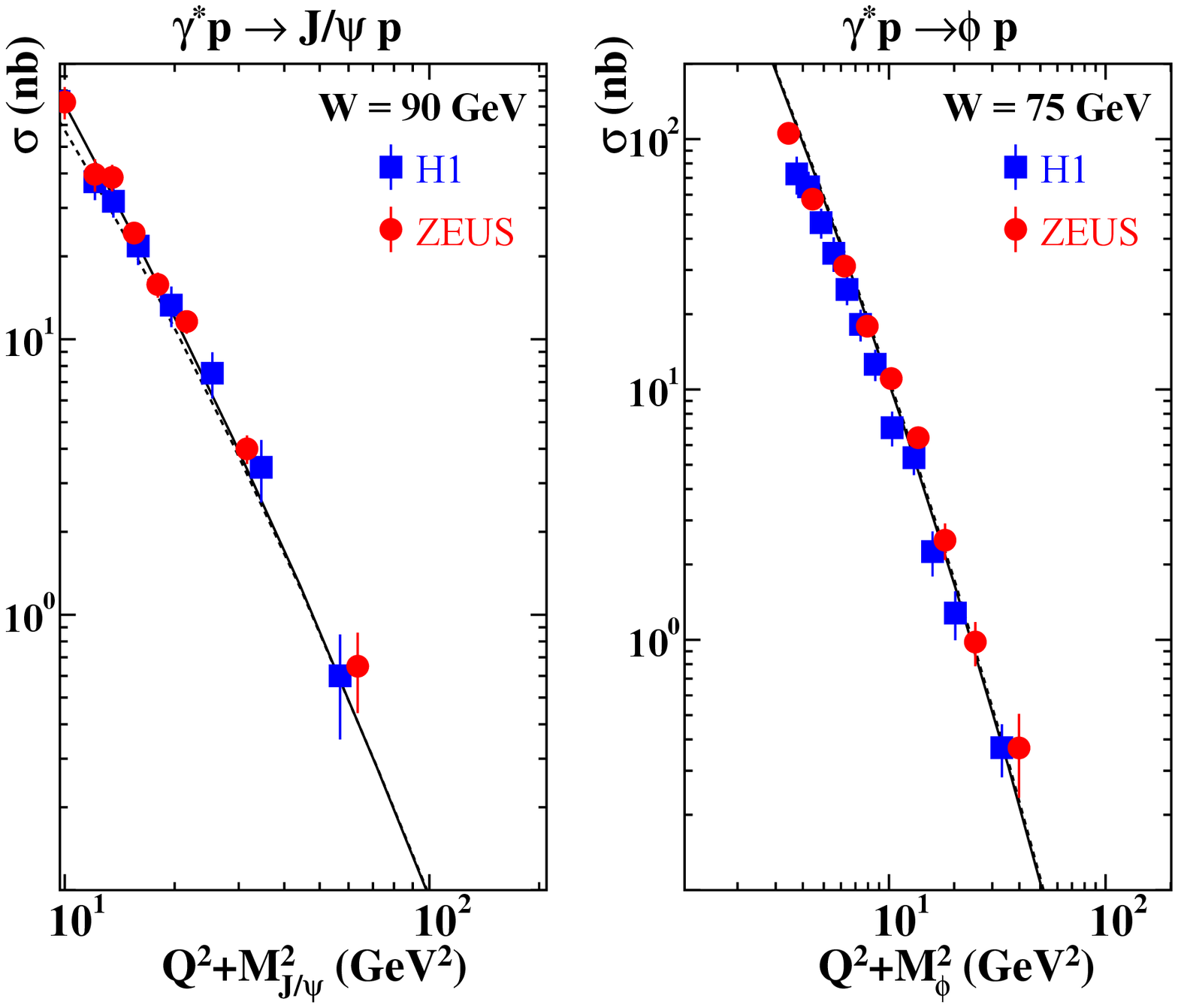}
\includegraphics[angle=270,width=0.49\textwidth,clip]{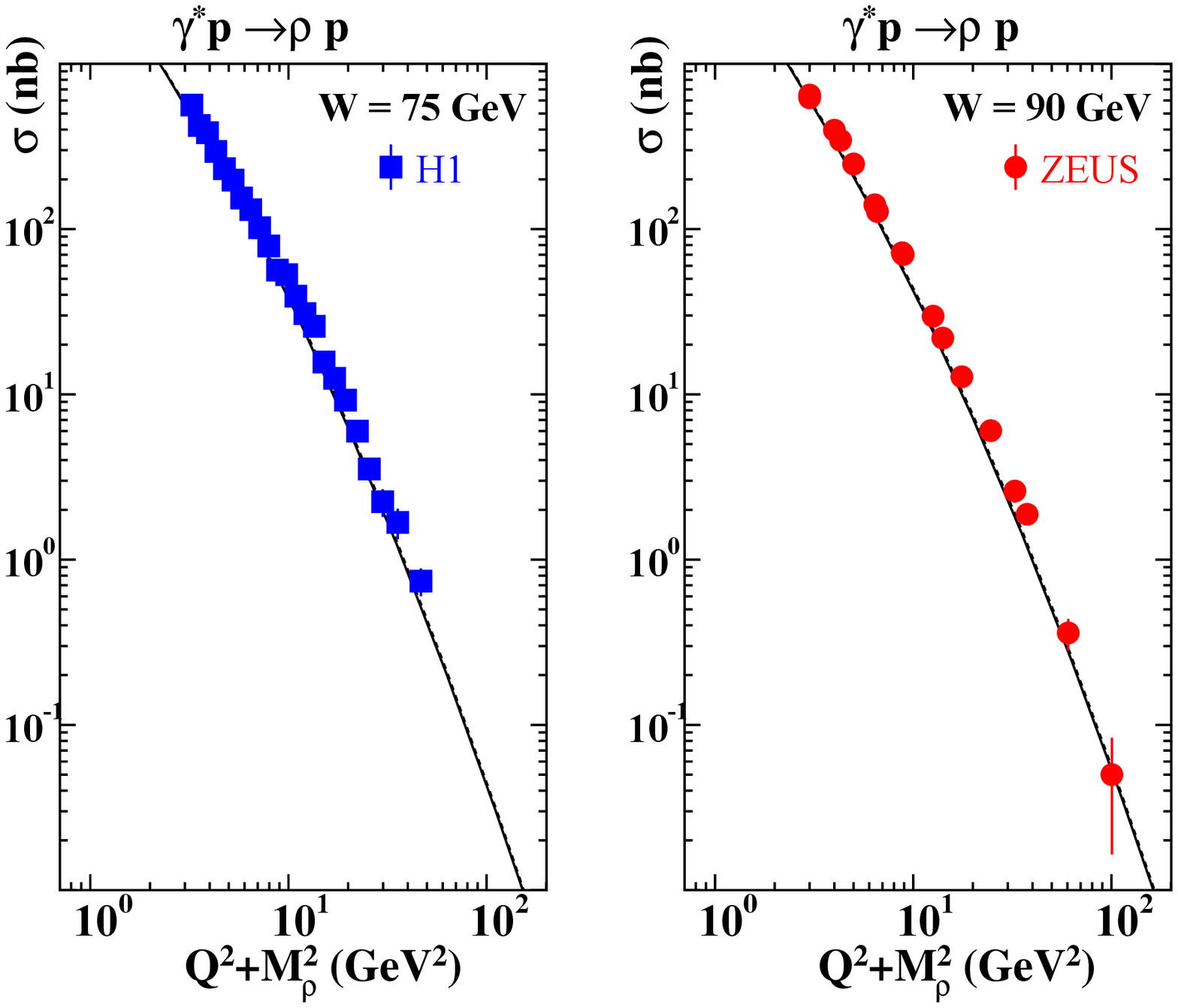}
  \caption{Total vector meson cross-section $\sigma$ for $J/\Psi$, $\phi$ and $\rho$ as a function of $Q^2+M^2_V$ compared to results from the IP-Sat model using the two parameter sets given in table \ref{t-1} with $m_c=1.27$ GeV (solid line) and $m_c=1.4$ GeV (dashed line).  The ZEUS data points are scaled to H1 $Q^2$-bins using the $Q^2$ dependence measured by ZEUS of the form $\sigma\propto (Q^2+M^2_V)^{-2.44}$. The experimental data are from H1 and ZEUS collaborations \cite{Chekanov:2002xi,Chekanov:2004mw,Aktas:2005xu,Chekanov:2005cqa,Aaron:2009xp,Chekanov:2007zr}.  }
  \label{f-vq}
\end{figure}

\begin{figure}
\includegraphics[angle=270, width=1 \textwidth,clip]{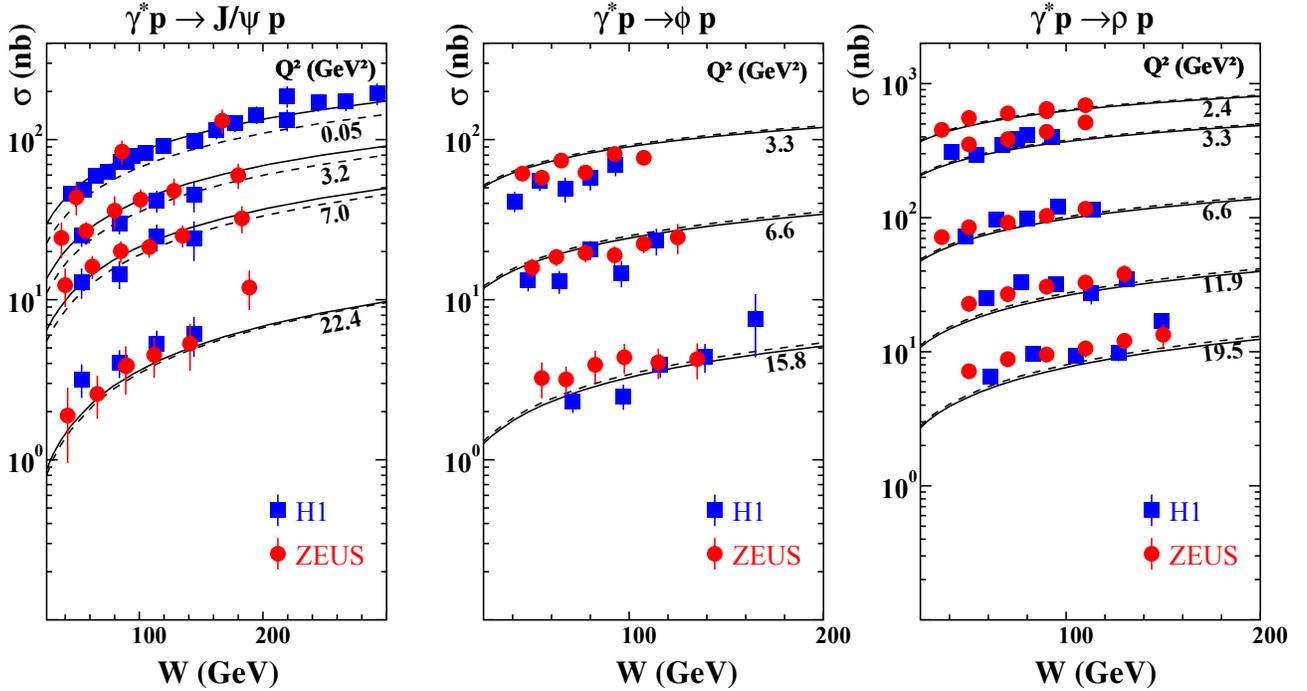}
  \caption{Total vector meson cross-section as a function of $W$ compared to results from the IP-Sat model using the two parameter sets given in table \ref{t-1} with $m_c=1.27$ GeV (solid line) and $m_c=1.4$ GeV (dashed line). As in \fig{f-vq}, the experimental data are from H1 and ZEUS collaborations \cite{Chekanov:2002xi,Chekanov:2004mw,Aktas:2005xu,Chekanov:2005cqa,Aaron:2009xp,Chekanov:2007zr}. }
  \label{f-vw}
\end{figure}

\begin{figure}
  \includegraphics[angle=270, width=1 \textwidth,clip]{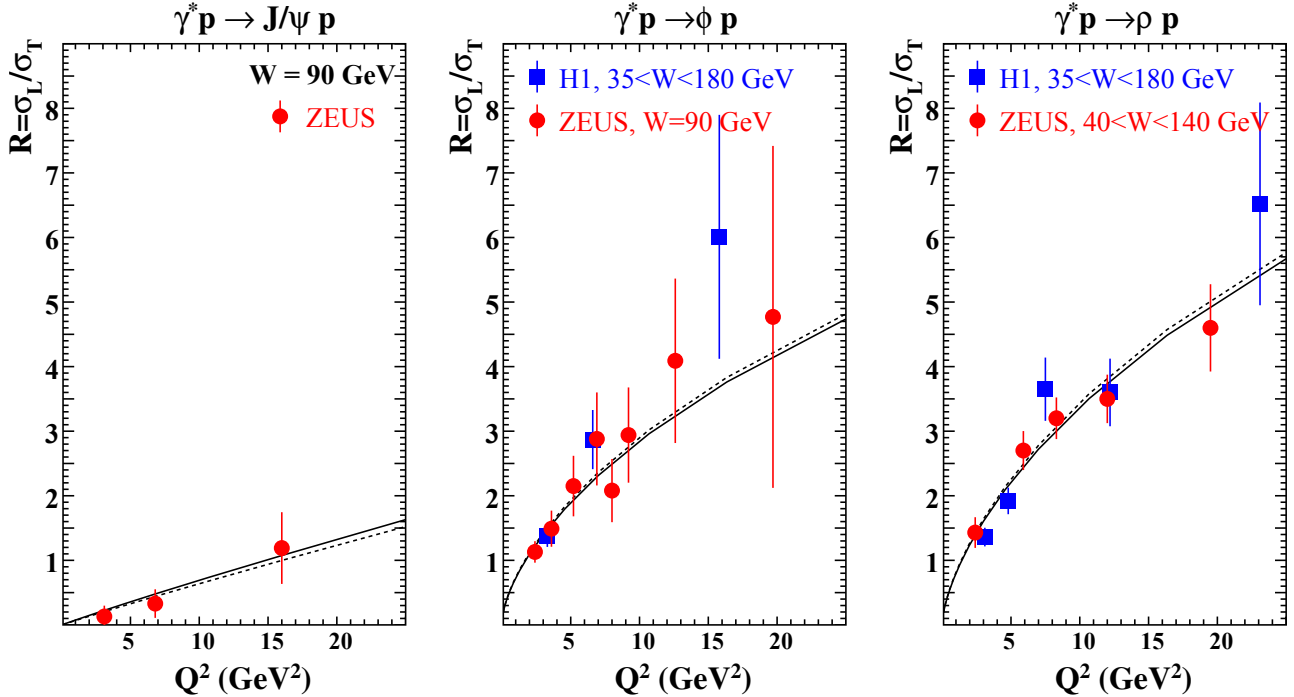}
  \caption{The ratio $R=\sigma_L/\sigma_T$ as a function of $Q^2$. The solid and dashed lines are results from the IP-Sat model with charm mass $m_c=1.27$ and $1.4$ GeV respectively.  The experimental data are from H1 and ZEUS collaborations \cite{Chekanov:2002xi,Chekanov:2004mw,Aktas:2005xu,Chekanov:2005cqa,Aaron:2009xp,Chekanov:2007zr}. }
  \label{f-rv}
\end{figure}

\begin{figure}
 % \centering
  \includegraphics[width=0.95\textwidth,clip]{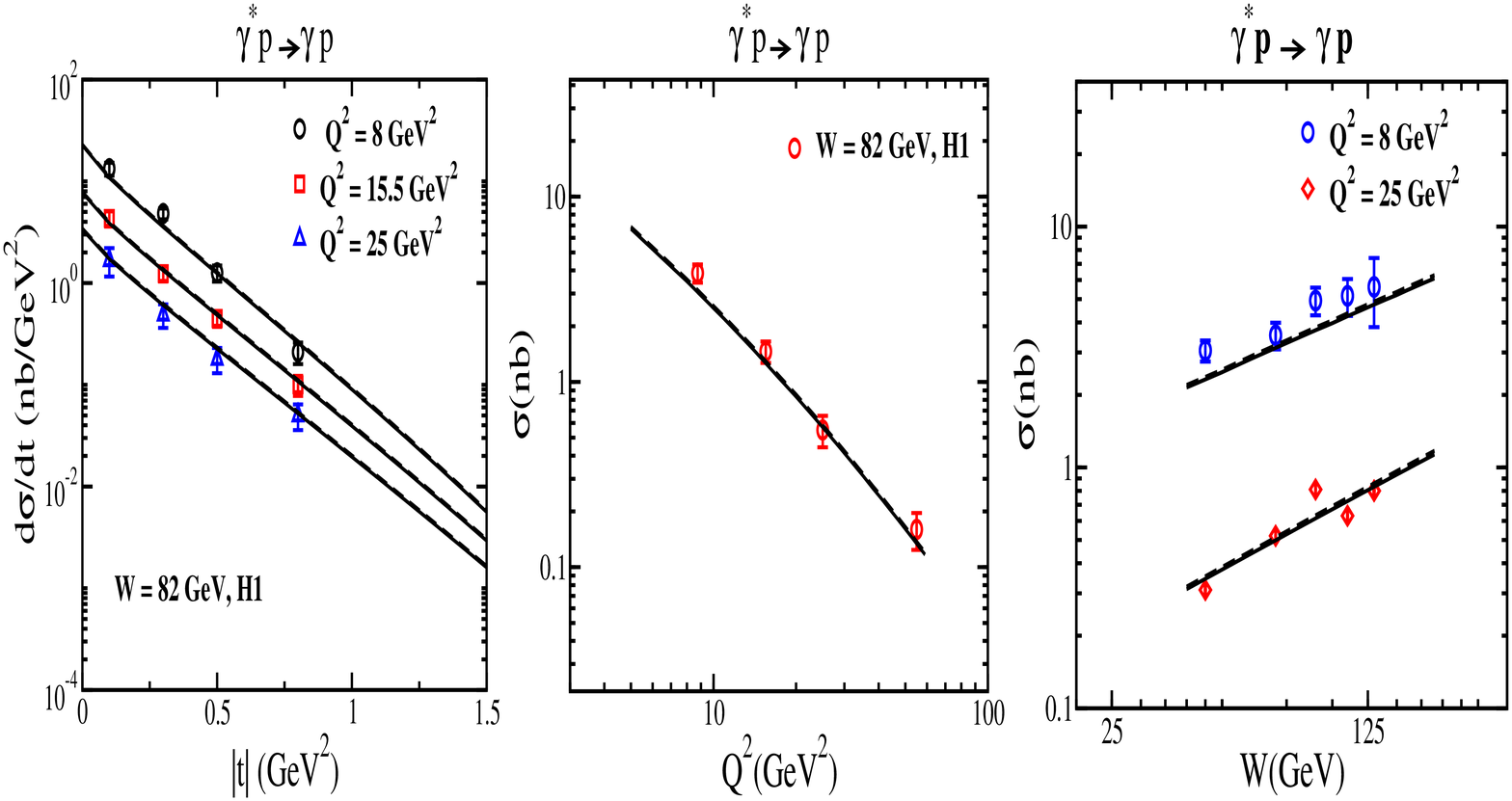}
  \caption{ Left: Differential DVCS cross-section as function of $|t|$. Middle: Total DVCS cross-section as a function of $Q^2$.  The solid and dashed lines are results from the IP-Sat model with charm mass $m_c=1.27$ and $1.4$ GeV respectively.  Right: Total DVCS cross-section as a function of $W$. The experimental data are from H1 and ZEUS collaborations \cite{Aaron:2009ac,Chekanov:2008vy}. }
  \label{f-dvcs}
\end{figure} 

To further test the IP-Sat dipole model, in Figs.\,(\ref{f-vt},\ref{f-vq},\ref{f-vw})  we confront experimental data from H1 and ZEUS with model predictions for the $|t|$, $Q^2$ and $W$-dependence of the vector mesons  $J/\Psi$, $\phi$ and $\rho$ production in various kinematics. Note that none of the data sets in those figures are used into our fits and the results of the model are fixed by $B_G$ alone. For the total cross-section, we performed the integral over $|t|$ up $1\,\text{GeV}^2$. 
 It is generally seen that the agreement between our results and  data is excellent.  In Figs.\,\ref{f-vt},\ref{f-vq} and \ref{f-vw} we show our results calculated with two dipole parameter sets given in table \ref{t-1} corresponding to charm mass $m_c=1.27$ GeV (solid lines) and $m_c=1.4$ GeV (dashed lines). Note that both set of parameters give very similar results. 
The $J/\Psi$ production cross-section is more sensitive to the charm quark mass at low $Q^2$ compared to light vector meson production cross-sections and the DVCS. This is due to the fact that the scale in the integrand of the cross-section is set by $\epsilon_f$ defined in \eq{eps}, and only at low virtualities for $Q^2<m_f^2$ does the cross-section become sensitive to quark mass.  It is seen from \fig{f-vw} that the $W$-dependence of the cross-section follows a power-law behavior of the form $\sigma\propto W^\delta$. Our extracted values of $\delta$ at different kinematics for various vector mesons are also in perfect agreement with experimental data from H1 and ZEUS \cite{Chekanov:2002xi,Chekanov:2004mw,Aktas:2005xu,Chekanov:2005cqa,Aaron:2009xp,Chekanov:2007zr} (for brevity, we do not show the comparison here).  

In \fig{f-rv}, we show the ratio of the longitudinal to the transverse cross-sections $R=\sigma_L/\sigma_T$ for $J/\Psi$, $\phi$ and $\rho$ production as a function of $Q^2$. The ratio $R$ provides vital information about the effective averaged dipole transverse-size which participates in the interactions. The transverse component of the cross-section is generally dominated by larger dipole sizes than the longitudinal component. This leads to a relative increase of the $t$-slope of the  transverse component compared to longitudinal one, i.e. $B_{DT}>B_{DL}$ seen also in data. Therefore one expects the ratio  $R=\sigma_L/\sigma_T$ to increase with $Q^2$, as can be seen in \fig{f-rv}. Note that the theory curves in  \fig{f-rv}  were calculated at a fixed $W=90$ GeV while the experimental data are for a range of $W$-bin shown in the plot. However, the $W$ dependence of the ratio is very weak for a fixed $Q^2$ \cite{Chekanov:2002xi,Chekanov:2004mw,Aktas:2005xu,Chekanov:2005cqa,Aaron:2009xp,Chekanov:2007zr}.

In \fig{f-dvcs}, we compare the IP-Sat results for  DVCS with the experimental data from H1: in the left panel, we show the the $t$-distribution at various values $Q^2$ and fixed $W=82$ GeV, in the middle panel we show  $\sigma$ as a function of $Q^2$ with fixed $W$, and in right panel the total cross-section $\sigma$ as a function of $W$ with different fixed values of $Q^2$. It is generally seen that the agreement with data is remarkably good. We do not compare with the ZEUS data, but H1 and ZEUS data for DVCS are consistent with each other \cite{Aaron:2009ac,Chekanov:2008vy} and our description of ZEUS data is of the same quality as for H1 data. The $W$-dependence of the cross-section in \fig{f-dvcs} can be described by a power-law behavior $W^\delta$ where $\delta\approx 0.76\pm 0.02$ at $Q^2=8\,\text{GeV}^2$ and $\delta\approx 0.83\pm 0.02$ at $Q^2=25\,\text{GeV}^2$ extracted from the IP-Sat model is in agreement with the experimental value $\delta=0.77\pm 0.23\pm 0.19$  \cite{Aaron:2009ac,Chekanov:2008vy}.

\section{Summary}
In this paper, we confronted the IP-Sat dipole model with the new combined data from HERA which are significantly more precise than previous data sets. The IP-Sat model has very few free parameters and they are fixed from a fit to the reduced cross-section $\sigma_r$ at small-$x$. Model results are then compared to $F_2, F_2^{c\bar{c}}, F_L$ and data from exclusive diffractive processes such as vector meson production and the DVCS  with the available data from HERA including the new combined data for $\sigma_r^{c\bar{c}}$. Overall, the model provides an excellent description of data in the range $Q^2\in [0.75, 650]\,\text{GeV}^2$ and $x\le 0.01$. 
The quality of this description relies on a $Q^2_{\text{min}}$ of $0.75$ GeV$^2$; one observes a steady increase in $\chi$-squared with decreasing values of $Q^2_{\text{min}}$. Other key features of this novel fit are the lower values for the light quark masses, and the four flavor running of the coupling. 

In addition to the inclusive data, we showed that most features of HERA data on exclusive processes are reproduced by the IP-Sat model. 
The $W$-dependence of the cross-section for vector meson production and the DVCS in Figs.\,(\ref{f-vw},\ref{f-dvcs}) shows that the energy-dependence of the dipole amplitude is correctly modeled, while the $|t|$-dependence in  Figs.\,(\ref{f-vt},\ref{f-dvcs}) suggest the shape of the proton is correctly implemented in our model. Finally, the $Q^2$-dependence of the cross-section in Figs.\,(\ref{f-vq},\ref{f-dvcs}) indicates that the virtuality dependence of the gluon distribution and its evolution with the dipole transverse-size are correctly modeled.  Moreover, our model also reproduces the experimental data for the ratio $R=\sigma_L/\sigma_T$ shown in \fig{f-rv}. This is another piece of evidence in favor of the dipole picture and the fact that the dipole transverse size sets the relevant length scale in interactions for both transverse and longitudinally polarized virtual photons.  We stress that the $t$-distribution of all three vector mesons ($J/\psi$, $\phi$, $\rho$) as well as DVCS  can be correctly reproduced by fixing  only one parameter $B_G$ despite the fact that the vector meson and DVCS wave functions are very different. This strongly hints at universality of the extracted impact-parameter distribution of the proton.   

Unintegrated gluon distributions can be extracted from the dipole amplitude in the IP-Sat model, and in a $k_\perp$ factorized formalism have been  compared to data from RHIC and the LHC in proton-proton and proton-nucleus collisions~\cite{Tribedy:2010ab,Tribedy:2011aa}.  In addition, the IP-Sat model is the basis of the IP-Glasma model~\cite{Schenke:2012wb} of initial conditions in 
heavy ion collisions. However, the parameters employed in these studies were determined from data from H1 and ZEUS predating the combined data sets for the proton. It will be interesting to see what the impact of the new fits are on final state observables in proton-proton, proton-nucleus and nucleus-nucleus collisions.

\begin{acknowledgments}
A.R and M.S thank Yuri Ivanov for technical support of  the USM HPC cluster. 
This work is supported in part by Fondecyt grant 1110781 and 1120920. M. van de K. and R. V. acknowledge partial support from an LDRD grant from Brookhaven Science Associates.
M. van de K.'s work was supported by scholarships of the VSB foundation. R. V.'s work is supported by DOE Grant DE-AC02-98CH10886.
\end{acknowledgments}

%\newpage

\newpage

 \end{document}